\documentclass[conference]{IEEEtran}

\usepackage{amsmath,amssymb,amsfonts}
\usepackage{algorithmic}
\usepackage{graphicx}
\usepackage{textcomp}
\usepackage{xcolor}
\usepackage{orcidlink}

\usepackage[utf8]{inputenc}
\usepackage[T1]{fontenc}
\usepackage{cleveref}
\usepackage{hyperref}
\hypersetup{hidelinks,unicode}

\usepackage[style=ieee, backend=biber, url=false, hyperref, maxnames=5, minnames=1, maxcitenames=2, mincitenames=1]{biblatex}

\DeclareFieldFormat{urldate}{(visited on \thefield{urlyear}\adddot\mkmonthzeros{\thefield{urlmonth}}\adddot\mkdayzeros{\thefield{urlday}}\isdot)}

\DeclareBibliographyDriver{online}{%
    \usebibmacro{bibindex}%
    \usebibmacro{begentry}%
    \usebibmacro{author/editor+others/translator+others}
    \setunit{\labelnamepunct}\newblock
    \usebibmacro{title}%
    \newunit
    \printlist{language}%
    \newunit\newblock
    \usebibmacro{byauthor}%
    \newunit\newblock
    \usebibmacro{byeditor+others}%
    \newunit\newblock
    \printfield{version}%
    \newunit
    \printfield{note}%
    \newunit\newblock
    \printlist{organization}
    \newunit\newblock
    \usebibmacro{date}%
    \newunit\newblock
    \iftoggle{bbx:eprint}
    {\usebibmacro{eprint}}
    {}%
    \newunit\newblock
    \usebibmacro{url+urldate}%
    \newunit\newblock
    \usebibmacro{addendum+pubstate}%
    \setunit{\bibpagerefpunct}\newblock
    \usebibmacro{pageref}%
    \usebibmacro{finentry}}

\usepackage{acronym}
\usepackage{dblfloatfix}
\usepackage{tabto}
\usepackage{enumerate}
\usepackage{siunitx}

\usepackage{multirow}
\usepackage{calc}
\usepackage{pifont}

\usepackage{tikz}
\usetikzlibrary{math,arrows.meta,positioning,calc,decorations.pathreplacing,calligraphy} 
\tikzmath{\bytewidth=0.5; \uintwidth=4*\bytewidth; \blockheight=0.8; \bitshoffset=-0.20; }

\definecolor{ecfviolet}{HTML}{ddd6e5}
\definecolor{ecfgreen}{HTML}{d7e3bf}
\definecolor{ecforange}{HTML}{fbd7bb}
\definecolor{ecfblue}{HTML}{b7dde8}
\definecolor{ecfpurple}{HTML}{f5d7f2}
\definecolor{ecfyellow}{HTML}{fee599}

\tikzset{
    u32le/.style={draw, rectangle, minimum width=\uintwidth cm, text width=\uintwidth cm, minimum height=\blockheight cm, inner sep=0pt, outer sep=0pt, align=center, font=\small}, 
    u8/.style={draw, rectangle, minimum width=\bytewidth cm, text width=\bytewidth cm, minimum height=\blockheight cm, inner sep=0pt, outer sep=0pt, align=center, font=\small}, 
    u8a/.style={draw, rectangle, minimum width=#1*\bytewidth cm, text width=#1*\bytewidth cm, minimum height=\blockheight cm, inner sep=0pt, outer sep=0pt, align=center, font=\small}, 
    half height/.style={minimum height=0.5*\blockheight cm},
}

\tikzset{
    pics/lock/.style n args={1}{
        code={
            \path [name=lock body, draw, fill, even odd rule]
            {[rounded corners=1mm] (0mm, 0mm) -- ++(0, 5mm) -- ++(5mm, 0) [rounded corners=2mm] -- ++(0, -5mm) -- cycle}
            {[sharp corners] (1.75mm, 1.25mm) -- ++(0.15mm, 1mm) arc (225:-45:0.75mm) -- ++(0.15mm, -1mm) -- cycle};
            \ifthenelse{\equal{#1}{open}}{
                \path [name=lock shackle, draw] (1mm, 5mm) -- ++(0, 2mm) arc (180:0:1.5mm) -- ++(0, -0.5mm);
            }{
                \path [name=lock shackle, draw] (1mm, 5mm) -- ++(0, 1mm) arc (180:0:1.5mm) -- ++(0, -0.5mm) -- ++(0, -1mm);
            }
        }
    }
}

\usepackage{nicematrix}
\usepackage{caption}
\usepackage{subcaption}
\usepackage{ifthen}

\DeclareMathOperator{\keyexchange}{Kex}
\DeclareMathOperator{\sign}{Sign}
\DeclareMathOperator{\verify}{Vrfy}
\DeclareMathOperator{\encrypt}{Enc}
\DeclareMathOperator{\decrypt}{Dec}
\DeclareMathOperator{\hash}{H}
\DeclareMathOperator{\gen}{Gen}
\DeclareMathOperator{\random}{Random}
\DeclareMathOperator{\convert}{Convert}
\DeclareMathOperator{\utf}{UTF8}
\DeclareMathOperator{\modify}{Modify}

\newcommand{\publickey}[2]{\textnormal{pk}_\textnormal{#1}^\textnormal{#2}}
\newcommand{\privatekey}[2]{\textnormal{sk}_\textnormal{#1}^\textnormal{#2}}
\newcommand{\sharedsecret}{\textnormal{ss}}
\newcommand{\symmetrickey}[1]{\textnormal{k}_\textnormal{#1}}
\newcommand{\convertSX}{\convert^{\textnormal{S} \rightarrow \textnormal{X}}}
\newcommand{\convertskpk}{\convert^{\textnormal{sk} \rightarrow \textnormal{pk}}}
\newcommand{\trunc}[2]{\left[#1, \dots, #2\right]}
\newcommand{\mfield}[1]{\textnormal{\field{#1}}}
\newcommand{\xor}{\oplus}

\acrodef{AEAD}{Authenticated Encryption with Associated Data}
\acrodef{AES}{Advanced Encryption Standard}
\acrodef{BLOB}{Binary Large Object}
\acrodef{CI}{Continuous Integration}
\acrodef{ECF}{Encrypted Container File}
\acrodef{GCM}{Galois/Counter Mode}
\acrodef{GPG}{GNU Privacy Guard}
\acrodef{KDF}{Key Derivation Function}
\acrodef{PGP}{Pretty Good Privacy}
\acrodef{PKI}{Public Key Infrastructure}
\acrodef{PoC}{Proof of Concept}

\newcommand{\gc}{\textit{git-crypt}}
\newcommand{\jak}{\textit{jak}}

\newcommand{\at}{\makeatletter@\makeatother}
\newcommand{\ecf}{\ac{ECF}}
\newcommand{\ecfs}{\acp{ECF}}

\newcommand{\yes}{\ding{51}}
\newcommand{\no}{\ding{55}}

\newcommand{\dt}[1]{\texttt{\small#1}}
\newcommand{\field}[1]{\textsc{\small\lowercase{#1}}}
\newcommand{\hex}[1]{\texttt{\small0x#1}}
\newcommand{\enquote}[1]{``#1''}
\newcommand{\fdt}[2]{\field{#1} of type \dt{#2}}
\newcommand{\csharp}[1]{\texttt{\small#1}} 

\newenvironment{fielditemize}{\itemize \setlength{\itemsep}{0.25em}}{\enditemize}

\crefname{subsection}{subsection}{subsections}
\Crefname{subsection}{Subsection}{Subsections}

\crefalias{ciphersuite}{enumerate}
\crefname{ciphersuite}{cipher suite}{cipher suites}
\Crefname{ciphersuite}{Cipher Suite}{Cipher Suites}

\makeatletter
\renewcommand{\fnum@figure}{Figure~\thefigure}
\makeatother

\usepackage{ifpdf}
\ifpdf
\pdfoutput=1 
\pdftrue
\fi

\makeatletter
\def\ps@IEEEtitlepagestyle{
    \def\@oddfoot{\mycopyrightnotice}
    \def\@evenfoot{}
}
\def\mycopyrightnotice{
    {\footnotesize
		\begin{minipage}{0.8\textwidth}
		  \centering
	   	Please cite as: Tobias J. Bauer and Andreas Aßmuth, ``Securing Confidential Data For Distributed Software Development Teams: Encrypted Container File,'' \emph{International Journal On Advances in Security}, vol.~17, no.~1 and 2, pp.~11--28, 2024.
		\end{minipage}
    }
}
\makeatother

\begin{document}
\title{\bfseries\Large Securing Confidential Data For Distributed Software Development Teams:\linebreak Encrypted Container File\vspace*{-4mm}}

\author{%
    \IEEEauthorblockN{Tobias J. Bauer\,\orcidlink{0009-0006-1073-3971}} 
    \IEEEauthorblockA{%
    Fraunhofer Institute for Applied and Integrated Security\\
    Weiden, Germany\\
    e-mail: {\tt tobias.bauer@aisec.fraunhofer.de}
    \vspace*{-8mm}
    }
    \and
    \IEEEauthorblockN{Andreas Aßmuth\,\orcidlink{0009-0002-2081-2455}}
    \IEEEauthorblockA{%
    Ostbayerische Technische Hochschule Amberg-Weiden\\
    Amberg, Germany\\
    e-mail: {\tt a.assmuth@oth-aw.de}
    \vspace*{-8mm}
    }
}

\maketitle

\begin{abstract}
In the context of modern software engineering, there is a trend towards Cloud-native software development involving international teams with members from all over the world. Cloud-based version management services like GitHub are commonly used for source code and other files. However, a challenge arises when developers from different companies or organizations share the platform, as sensitive data should be encrypted to restrict access to certain developers only. This paper discusses existing tools addressing this issue, highlighting their shortcomings. The authors propose their own solution, Encrypted Container Files, designed to overcome the deficiencies observed in other tools.
\end{abstract}

\renewcommand\IEEEkeywordsname{Keywords}

\section{Introduction}\label{sec:introduction}
Modern software development processes are agile, span organizational boundaries, and are carried out by teams whose members are not confined to one location. They produce Cloud-native software components that communicate with each other. However, there is strong interest in ensuring the confidentiality of the transmitted data. Usually, encryption algorithms fulfill this task, which is also true for storing confidential data, e.g., in a database. Often during the development of Cloud-native software components, passwords, private certificate keys, and symmetric encryption keys arise. On the one hand, they must be kept secret, but on the other hand, they must be managed in a sensible way and some authorized team members have to have access to these secrets. In the following, we refer to these types of data as \textit{confidential data}. To address the emerging conflict of these two opposing objectives, this paper aims to extend our solution we presented in~\autocite{Bauer2023}.

The practice of Continuous Integration (cf.~\autocite{Fowler2006}) is a long-lasting trend in software development~\autocite{Intellectsoft2024}\autocite{Netsolutions2023}. The developed applications are then distributed and deployed automatically via Continuous Delivery and Continuous Deployment~\autocite{Intellectsoft2024}. However, designing a secure deployment process or pipeline is difficult and requires special tooling as hinted in~\autocite{Bajpai2022}\autocite{Cybarark2023}. As a positive aspect, the awareness to build these automated processes in a secure way has grown over the past two years (cf.~\autocite{Netsolutions2023}\autocite{Samsolutions2024}). This leads to incorporating the security aspect into the development and operation or \textit{DevOps} cycle and stressing this new component by naming the new approach \textit{DevSecOps}~\autocite{Netsolutions2023}. Although best practice guides stress the need for securing confidential data, e.g., in~\autocite{Cybarark2023} and~\autocite{Thenewstack2022}, they lack providing a pipeline vendor-independent solution (cf.~\autocite{Cybarark2023}\autocite{Devsecopsguides2024}) or advertise third party (cloud) services (cf.~\autocite{Thenewstack2022}). Some implementations have been found to expose secrets~\autocite{Koishybayev2022}.

Software developing teams often use version control systems, e.g., \textit{git}~\autocite{Git2022}, to manage both the source code and other artifacts. The aforementioned practice of Continuous Integration requires the version control system to contain \textit{all} artifacts, which -- by definition -- includes secrets we described earlier. However, storing confidential data without protective measures against unauthorized access would be grossly negligent and poses a serious hazard. There are tools available that offer the encryption of such confidential data before storing it within the version control system. However, we observed multiple drawbacks and shortcomings when using these tools. We present \ecf, our solution for a Cloud-based, hybrid-encrypted data storage aimed at software development teams. With \ecf\ we extend the functionality of currently available tools and simultaneously overcome the found deficiencies.

The remainder of this paper is structured as follows: in \Cref{sec:relatedwork}, we take a look at three existing tools and discuss their benefits and shortcomings. Next, we provide a high-level workflow description of the usage of \ecf\ in \Cref{sec:app}. This leads to the requirements and the file format description presented in \Cref{sec:requirements}. In \Cref{sec:ops}, we describe the operations of an \ecf\ in detail. Since our initial presentation of \ecf, we extended our work in various ways, which are described in \Cref{sec:improvements}. Finally, we conclude this paper in \Cref{sec:conclusion}.

\section{Related Work}\label{sec:relatedwork}
Whilst researching available solutions for the problem described above, we came across two projects: \jak\ and \gc. In our opinion, however, both have some disadvantages, which we will briefly describe below. Lastly, we argue that \ac{PGP} or \ac{GPG} can solve the problem but the usage is inconvenient and error prone.

\subsection{\jak}\label{sec:relatedwork:jak}
\jak~\autocite{Dispel2017}, a tool written in Python, allows the encryption of files with symmetric keys using the \ac{AES}. With \jak, keys can be generated and stored in a key file, which itself is not encrypted. If a special text file containing a list of the file names of the corresponding files is added to the repository, \jak\ can perform automatic encryption and decryption with a single command. As is well known, the so-called key exchange problem cannot be solved using encryption with symmetric keys only. This problem also exists with \jak, which relies exclusively on \ac{AES} as stated before. The tool therefore cannot offer a solution to the key exchange problem, but leaves it up to the developers/users to take care of it themselves. The distribution of confidential data (keys) becomes increasingly challenging, especially as the size of the development team increases, which is why we believe that the practical use of \jak\ is limited. Another issue with \jak\ is that the content of confidential files is stored unencrypted on the developers' computers. The reason for this is that \jak\ decrypts these files during checkout and re-encrypts them before committing. A prerequisite for the secure use of \jak\ is therefore that external parties may only have read-only access to the repository. In addition, third parties must be prevented from accessing the developers' computers. If an attacker manages to gain access to such a computer, they can read the confidential information.\par 
For further information on \jak, please refer to the project documentation~\autocite{Dispel2017}.

\subsection{\gc}\label{sec:relatedwork:git-crypt}
\gc~\autocite{Ayer2022} allows the encryption of files within a \textit{git} repository with \ac{AES} as well. This results in the same restrictions as with \jak\ with regard to the necessary access restrictions for third parties. However, \gc\ offers a solution to the key exchange problem by using \ac{GPG}. For this purpose, the public \ac{GPG} keys of the recipients are added to the repository. When encrypting confidential files within the repository, \gc\ generates an asymmetrically encrypted key file for each recipient. Each authorized recipient can use this to access the symmetric key, which enables each of these recipients to decrypt the confidential files in this repository. \gc\ is implemented in a way that all confidential files are encrypted with the same symmetric key and this key must therefore be communicated to all recipients that are added to the repository. Unfortunately, this means that it is not possible to differentiate between different authorized recipients when accessing different confidential files. In our opinion, such a feature would be absolutely desirable. For example, confidential information about the production environment should only be accessible by the production team. However, there may also be other confidential files to which not only (some members of) the production team can have access. But this cannot be realized with \gc. As with \jak, the contents of confidential files that are stored locally on the developer's computer are not secured. The reason for this is that \gc\ decrypts the confidential data during checkout (same as with \jak). The integration of \gc\ into the mechanisms of \textit{git} is optional, but recommended. With \gc, we also miss an option of being able to remove authorized users from the confidential files at a later date, for example if one of these developers leaves the project. In such a case, he or she should no longer have access to the confidential files. \autocite{Ayer2022}~justifies this by arguing that when using a version control system, a remote recipient can still access old versions of the repository and thus the confidential data stored in it. This is true, of course. However, it should also be borne in mind that a former developer should not have access to new certificates, passwords, etc. once the old ones have expired. Against the background of such changes of personnel, we consider this feature to be desirable. In addition, we recommend also changing the symmetric keys in this scenario.

\subsection{\acl{PGP}}\label{sec:relatedwork:pgp}
Another way to solve the problem would be to use \ac{PGP} or GPG directly. To do this, one of the developers would generate a random session key to encrypt the confidential data. The random session key would then be encrypted with the public key of another developer. The file consisting of the encrypted session key and the symmetrically encrypted ciphertext could then be stored in the online repository. Only the aforementioned second developer would be able to access the confidential data.\par 
As this brief description of \ac{PGP}/\ac{GPG} suggests, this solution does not scale very well. If the confidential data is required to be shared with multiple recipients, one such file per recipient must be created and stored in the repository. Of course, this is feasible in principle. However, it is important to note that each recipient must know which of these files is intended for them. In addition, it is absolutely impure that the same confidential data has to be stored in encrypted form in several instances. Changes to the content of the confidential data naturally require all these recipient-specific files to be recreated and stored. This makes this solution very challenging to use, and there is also the problem of keeping the content of all the confidential data's ciphertext files consistent. Just imagine that the confidential data of a project is spread over multiple files and each for different groups of recipients. Although handling with this solution is feasible in principle, it demands a great deal from everyone involved and is anything but convenient.

\section{Application}\label{sec:app}
In this section, we present a high-level description of multiple parties using an \ecf. In order to illustrate the basic operations, we follow a common workflow involving various participants representing the different parties. The generic users Alice, Bob, and Charlie use a command line-based tool to interact with an \ecf. This tool is also available as a part of this paper's accompanying source code\footnote{\url{https://github.com/Hirnmoder/ECF}}. Furthermore, all basic operations are depicted in \Cref{fig:app:overview:a,fig:app:overview:b}. Each subfigure corresponds to its analogously named subsection, e.g., \Cref{fig:app:create} visualizes \Cref{sec:app:create}.

This use case-oriented view serves as a foundation for the distillation of requirements of the \ecf\ format presented in \Cref{sec:requirements}. We describe the technical details in \Cref{sec:ops}, which follows the workflow presented in this section.

\subsection{Creating an \ecf}\label{sec:app:create}
Alice is the head of an operations team working for a company. Her task is to securely store secrets, such as server certificates for the company's website. Furthermore, a subset of her coworkers as well as an automated deployment job should get access to these secrets. Third parties, i.e., all other coworkers as well as any other person, must not get access to the sensitive data. Alice therefore chooses the \ecf\ format to protect and store the secret data.

As a first step, Alice wants to create an \ecf\ using the aforementioned command line tool (hereinafter referred to as the tool). To do so, Alice needs to generate a key pair consisting of a private and a public key. She may use this key pair for as many different \ecfs\ as she likes, but she can also generate as many different key pairs as she likes. In order to keep this workflow description basic, we assume that Alice generates a single key pair using the tool. Since the private key must not be accessible to any person other than Alice, storing it unencrypted is not an option. This is why the tool asks for a self-chosen password which is then used to encrypt Alice's private key before storing it to disk.

In \Cref{fig:app:create} and all other subfigures in \Cref{fig:app:overview:a,fig:app:overview:b}, we depict a key pair as a keyring with two keys attached. Additionally, we always use the same color for a user as well as their keys and attributes, which is red for Alice.

After this initial step, Alice uses the tool to create an empty \ecf. The tool then asks for her self-chosen password to get access to her private and public key. This is needed in order to add Alice as a recipient to the newly created \ecf\ and to store its initial content, i.e., the secret server certificate, inside. To indicate that Alice is a recipient of that newly created \ecf, we add an appropriately colored rectangle to it in \Cref{fig:app:create}.

When using a version control system, such as \textit{git}~\autocite{Git2022}, Alice may now add the \ecf\ to the repository, commit the changes, and push it to a remote server. Since the sensitive data is encrypted within the \ecf\ and no recipient but Alice was added to the \ecf, only Alice is able to decrypt and access the secrets stored in that \ecf.

\subsection{Decrypting and Using an \ecf}\label{sec:app:use}
In order to access the data stored inside an \ecf, Alice needs to be a recipient of the \ecf\ in question. According to our workflow, Alice was added as a recipient to the \ecf\ during its creation. She, thus, can decrypt and view the content of the \ecf. In addition to that, Alice is also able to alter the recipient set as well as the sensitive data. Both of these actions are described later in \Cref{sec:app:add,sec:app:remove,sec:app:content}. This subsection is focused solely on the decryption and usage of an \ecf.

Alice uses the tool to access the encrypted content. The tool first asks for Alice's self-chosen password and then decrypts the \ecf\ using her private key. It then displays the decrypted content of the file for Alice to view and use. This operation is visualized in \Cref{fig:app:use}.

Usually, manual inspection of the content of an \ecf\ is rare. It is much more likely that other processes, e.g., an automated deployment job, needs to use the secrets and therefore needs to access them. Such a process should run in a trusted environment, e.g., on a self-hosted deployment server. For the \ecf, the process looks like any other recipient and, thus, Alice has to first add the process as a recipient to the \ecf\ in order for it to be able to access the stored secret data. The procedure for adding a recipient is described in \Cref{sec:app:add}. The process then can read (and possibly modify) the encrypted secrets inside the \ecf\ and perform its task, e.g., deploying the company's website with the server certificate.

\newcommand{\person}[3]{
    \coordinate (start) at #1;
    \fill[fill=#2] (start) -- ++(0, 0.855) circle (0.18cm);
    \fill[fill=#2] (start) -- ++(0, 0.45) circle (0.225cm);
    \fill[fill=#2] (start) -- ++(0.225, 0) -- ++(0, 0.45) -- ++(-0.45, 0) -- ++(0, -0.45) -- cycle;
    \node[anchor=north] at (start) {#3};
}

\newcommand{\confdata}[3][confidential\\data]{
    \coordinate (start) at #2;
    \draw (start) -- ++(0.25, 0) -- ++(0, 0.6) -- ++(-0.5, 0) -- ++(0, -0.6) -- cycle;
    \draw (start) ++(0, 0.3) node{#3};
    \node[anchor=north, text width=1.5cm, text centered] at (start) {#1};
}

\newcommand{\ecfpic}[2]{
    \coordinate (start) at #1;
    \draw (start) -- ++(0.25, 0) -- ++(0, 0.6) -- ++(-0.5, 0) -- ++(0, -0.6) -- cycle;
    \draw (start) ++(0, 0.3) node{\includegraphics[width=0.3cm]{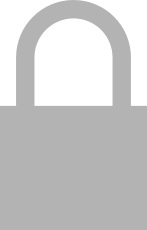}};
    \draw (start) ++(0, 0.3) node{#2};
    \node[anchor=north] at (start) {ECF};
}

\begin{figure}[b]
    \vspace*{-8mm}
    \begin{subfigure}[b]{\linewidth}
    \centering
    \begin{tikzpicture}
        \footnotesize
        \useasboundingbox (0, 0) rectangle (7.25, 2);
        \person{(0.5, 0.4)}{red}{Alice}
        \confdata{(1.75, 0.75)}{$x$}
        \draw[-latex] (2.75, 1.15) -- node[above]{create} (4.25, 1.15);
        \node[anchor=north] at (3.5, 1.15) {\includegraphics[height=0.5cm]{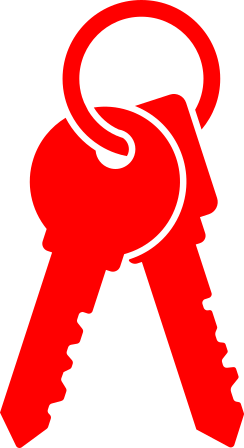}};
        \ecfpic{(5.25, 0.75)}{$x$}
        \draw[very thick, red] (5.4, 1.25) -- ++(0.2, 0); 
    \end{tikzpicture}
    \caption{\textcolor{red}{Alice} creates an \acl{ECF} (cf.~\Cref{sec:app:create}).}
    \label{fig:app:create}
    \end{subfigure}
    \begin{subfigure}[b]{\linewidth}
    \centering
    \begin{tikzpicture}
        \footnotesize
        \useasboundingbox (0, 0) rectangle (7.25, 2);
        \person{(0.5, 0.4)}{red}{Alice}
        \ecfpic{(1.75, 0.75)}{$x$}
        \draw[very thick, red] (1.9, 1.25) -- ++(0.2, 0); 
        \draw[-latex] (2.75, 1.15) -- node[above]{decrypt} (4.25, 1.15);
        \node[anchor=north] at (3.5, 1.15) {\includegraphics[height=0.5cm]{keyring_alice.png}};
        \confdata{(5.25, 0.75)}{$x$}
    \end{tikzpicture}
    \caption{\textcolor{red}{Alice} decrypts and uses an \acl{ECF} (cf.~\Cref{sec:app:use}).}
    \label{fig:app:use}
    \end{subfigure}
    \begin{subfigure}[b]{\linewidth}
    \centering
        \begin{tikzpicture}
        \footnotesize
        \useasboundingbox (0, 0) rectangle (7.25, 2);
        \person{(0.5, 0.4)}{red}{Alice}
        \ecfpic{(1.75, 0.75)}{$x$}
        \draw[very thick, red] (1.9, 1.25) -- ++(0.2, 0); 
        \node[anchor=south] at (2.66, 0.65) {\includegraphics[height=0.5cm]{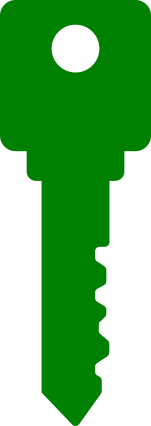}};
        \node[anchor=south] at (3.00, 0.65) {\includegraphics[height=0.5cm]{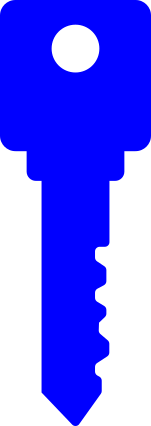}};
        \node[anchor=south] at (3.33, 0.65) {\includegraphics[height=0.5cm]{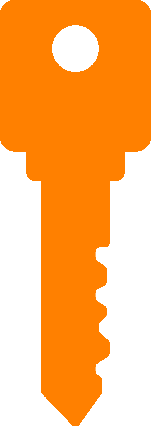}};
        \node[anchor=north, text width=1.5cm, text centered] at (3, 0.7) {public\\keys};
        \draw[-latex] (4, 1.15) -- node[above]{add} (5.5, 1.15);
        \node[anchor=north] at (4.75, 1.15) {\includegraphics[height=0.5cm]{keyring_alice.png}};
        \ecfpic{(6.5, 0.75)}{$x$}
        \draw[very thick, red] (6.65, 1.25) -- ++(0.2, 0); 
        \draw[very thick, green!50!black] (6.65, 1.15) -- ++(0.2, 0); 
        \draw[very thick, blue] (6.65, 1.05) -- ++(0.2, 0); 
        \draw[very thick, orange] (6.65, 0.95) -- ++(0.2, 0); 
    \end{tikzpicture}
    \caption{\textcolor{red}{Alice} adds recipients \textcolor{green!50!black}{Bob}, \textcolor{blue}{Charlie} and the\\\textcolor{orange}{automated deployment job} (cf.~\Cref{sec:app:add}).}
    \label{fig:app:add}
    \end{subfigure}
    \caption{Overview of \ecf\ operations according to\\\Cref{sec:app:create,sec:app:use,sec:app:add}.}
    \label{fig:app:overview:a}
\end{figure}

\subsection{Adding Recipients}\label{sec:app:add}
We now assume the following situation: Alice has two coworkers that also need access to the secrets, e.g., as a substitute when she is ill. Furthermore, the aforementioned automated deployment job also needs access to the secrets in order to fulfill its task.

As a preparatory step, her coworkers Bob and Charlie need to generate a key pair each, analogous to what Alice did in \Cref{sec:app:create}. Then, Bob and Charlie export their public keys using the tool and sending those to Alice, e.g., via email. Bob and Charlie can perform these steps completely independently of each other and may use different computers and operating systems. Since Bob and Charlie keep their private keys private and only share their public keys, we use a single-key symbol for them in \Cref{fig:app:add}.

Since the automated deployment job is a process and not a person, Alice (or someone else she trusts for that matter) can generate a key pair for the process. This key pair can now be used to export the public key using the tool. The private key must be made available to the process while still keeping it safe from unauthorized access. This can be done by restricting access to the server the process is running on. This is also why the process should run within a trusted environment as stated in \Cref{sec:app:use}.

Still, nobody but Alice has currently access to the content stored in the \ecf\ Alice created in \Cref{sec:app:create}. To change this, Alice now receives and checks the exported public keys of the recipients-to-be. She has to make sure that the received public keys were not tampered with during the transmission. A way to ensure authenticity is to compare, e.g., fingerprints of these public keys via a second channel. For example, we assume that Alice and Bob work in the same facility. They now can verify that the transmission of Bob's public key did not alter it by displaying the fingerprints on their laptop screens and comparing it visually.

After verifying the received public keys, Alice can now add Bob, Charlie, and the automated deployment job as recipients to the \ecf. She uses the tool for each recipient addition. Since modifying the \ecf\ also involves decrypting it, Alice is prompted for her self-chosen password by the tool in order to access her private key. The order in which Alice adds the recipients does not matter. Afterwards, she may delete the exported public keys since they are no longer of use.

As a last step Alice must make the updated \ecf\ available to the other users. She may commit the changes and push them to a remote server using \textit{git}. Bob, Charlie, and the automated deployment job now gained access to the \ecf\ using their respective private keys and, thus, can perform the same actions Alice can, i.e., decrypting and using the \ecf, add or remove recipients, and change the encrypted content. In \Cref{fig:app:add}, we indicate that Bob, Charlie and the automated deployment job are now recipients by appropriately colored rectangles.

\subsection{Removing Recipients}\label{sec:app:remove}
So far, we have tackled the recipient addition procedure. We now assume that Bob leaves the company and should, therefore, no longer have access to the content of the \ecf. In this case, both, Alice and Charlie, can remove Bob from the recipient set -- we opted for Alice to perform this step as she is the head of the team. This operation is also visualized in \Cref{fig:app:remove}. Alice uses the tool to identify and remove Bob from the recipient set of the \ecf. As with any operation, the tool asks for her self-chosen password beforehand. Alice then commits the changes and pushes them to a remote server.

Bob now looses access to the current and future versions of the \ecf. However, Bob still has access to older versions. It is futile to withdraw access to older versions because (a~malicious) Bob could have copied the secret content anyways when he still had access to. An \ecf\ is designed to securely store secrets, such as access keys, private certificate keys, passwords, asymmetric and symmetric encryption keys, etc. As it is a good practice to rotate keys as described in~\autocite{Barker2020}, all secrets Bob once had access to will eventually expire. This is especially true for short-lived TLS certificates.

Nevertheless, it is advisable to only add recipients to an \ecf\ if it is really necessary in the first place~\autocite{Saltzer1975}. In fact, since each \ecf\ is independent of other \ecfs, the format allows for different recipient sets per file. Let us assume that Alice has two different secrets to share with different groups of people. Instead of creating a single \ecf\ and adding all recipients to it, Alice should rather create two \ecfs, one for each secret and only add the respective intended recipients to each.

\subsection{Modifying the Content}\label{sec:app:content}
The original \ecf\ Alice created contains the server certificates for the company's website. Since these certificates are usually short-lived, they have to be renewed regularly. Alice's coworker Charlie -- who got access to the \ecf\ as described in \Cref{sec:app:add} -- now has the task of modifying the content of the \ecf, which is depicted in \Cref{fig:app:content}.

Charlie uses the tool to modify the \ecf\ and therefore is prompted to enter her self-chosen password. Then she can view and edit the content of the \ecf. In order to fulfill her task, she removes the expired server certificate from the \ecf\ and inserts the renewed one. In \Cref{fig:app:content}, we label the original content as $x$ and the modified one as $y$. Last, she commits her changes using \textit{git} and pushes them to a remote server to make the new version of that \ecf\ available to all other recipients.

Now Alice, Charlie, and the automated deployment job can access the renewed certificate stored inside the \ecf. Bob, on the other hand, cannot decrypt this \ecf\ anymore as he was removed from the recipient set in \Cref{sec:app:remove}. 

\begin{figure}[t]
\ContinuedFloat
    \vspace*{-4mm}
    \begin{subfigure}[b]{\linewidth}
    \centering
    \begin{tikzpicture}
        \footnotesize
        \useasboundingbox (0, 0) rectangle (7.25, 2);
        \person{(0.5, 0.4)}{red}{Alice}
        \ecfpic{(1.75, 0.75)}{$x$}
        \draw[very thick, red] (1.9, 1.25) -- ++(0.2, 0); 
        \draw[very thick, green!50!black] (1.9, 1.15) -- ++(0.2, 0); 
        \draw[very thick, blue] (1.9, 1.05) -- ++(0.2, 0); 
        \draw[very thick, orange] (1.9, 0.95) -- ++(0.2, 0); 
        \draw[-latex] (2.75, 1.15) -- node[above]{remove} (4.25, 1.15);
        \node[anchor=north] at (3.5, 1.15) {\includegraphics[height=0.5cm]{keyring_alice.png}};
        \ecfpic{(5.25, 0.75)}{$x$}
        \draw[very thick, red] (5.4, 1.25) -- ++(0.2, 0); 
        \draw[very thick, blue] (5.4, 1.15) -- ++(0.2, 0); 
        \draw[very thick, orange] (5.4, 1.05) -- ++(0.2, 0); 
    \end{tikzpicture}
    \caption{\textcolor{red}{Alice} removes recipient \textcolor{green!50!black}{Bob} (cf.~\Cref{sec:app:remove}).}
    \label{fig:app:remove}
    \end{subfigure}
    \begin{subfigure}[b]{\linewidth}
    \centering
    \begin{tikzpicture}
        \footnotesize
        \useasboundingbox (0, -0.2) rectangle (7.25, 2);
        \person{(0.5, 0.4)}{blue}{Charlie}
        \ecfpic{(1.75, 0.75)}{$x$}
        \draw[very thick, red] (1.9, 1.25) -- ++(0.2, 0); 
        \draw[very thick, blue] (1.9, 1.15) -- ++(0.2, 0); 
        \draw[very thick, orange] (1.9, 1.05) -- ++(0.2, 0); 
        \confdata[modified confidential data]{(3, 0.75)}{$y$}
        \draw[-latex] (4, 1.15) -- node[above]{modify} (5.5, 1.15);
        \node[anchor=north] at (4.75, 1.15) {\includegraphics[height=0.5cm]{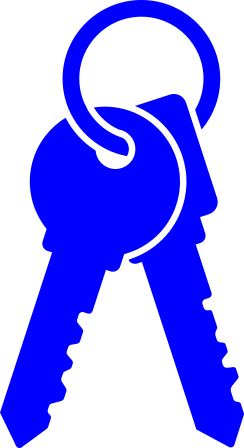}};
        \ecfpic{(6.5, 0.75)}{$y$}
        \draw[very thick, red] (6.65, 1.25) -- ++(0.2, 0); 
        \draw[very thick, blue] (6.65, 1.15) -- ++(0.2, 0); 
        \draw[very thick, orange] (6.65, 1.05) -- ++(0.2, 0); 
    \end{tikzpicture}
    \caption{\textcolor{blue}{Charlie} modifies the content (cf.~\Cref{sec:app:content}).}
    \label{fig:app:content}
    \end{subfigure}
    \caption{\textit{(Continued.)} Overview of \ecf\ operations according to\\\Cref{sec:app:remove,sec:app:content}.}
    \label{fig:app:overview:b}
\end{figure}

\section{Requirements Engineering and\\Internal File Structure}\label{sec:requirements}
Based on the workflow described in \Cref{sec:app}, we define requirements for the \ecf\ format. Then, we use these requirements to formulate design goals for the definition of the file format. In the following subsections, we describe the file structure and used data types in detail. A technical description of the operations on the \ecf\ format is given in \Cref{sec:ops}.

\subsection{Use Case Requirements}\label{sec:requirements:req}
We derive some requirements from the high-level workflow description given earlier. Furthermore, we take the features and weaknesses of the \jak\ and \gc\ tools as well as the plain-\ac{PGP} use into consideration. We have distilled the following requirements for the \ecf\ format, cf.~\autocite{Bauer2023}:

\begin{enumerate}
    \item\label{e:req:enc} The encryption of confidential data is mandatory,
    \item\label{e:req:mod} possibility to modify the confidential data,\\i.e., the content is writable,
    \item\label{e:req:key} key distribution is no prerequisite,
    \item\label{e:req:dec} decryption occurs not during checkout but on demand,
    \item\label{e:req:mul} support for many recipients,
    \item\label{e:req:arr} addition and removal of recipients,
    \item\label{e:req:mig} minimal information gain for external parties,
    \item\label{e:req:set} a customizable set of recipients per file, and
    \item\label{e:req:pfm} be reasonably fast to be used in production.
\end{enumerate}

\subsection{Design Goals}\label{sec:requirements:goals}
The requirements in mind, we have decided to define the following design goals for our \ecf\ format, cf.~\autocite{Bauer2023}. The numbers in brackets refer to the enumeration in \Cref{sec:requirements:req}.

\begin{itemize}
    \item Use of hybrid encryption (\Cref{e:req:enc,e:req:key,e:req:mul}),
    \item inclusion of recipient information to allow re-encryption on changes (\Cref{e:req:mod,e:req:mul,e:req:arr,e:req:set}),
    \item minimal information about recipients publicly available (\Cref{e:req:mig}),
    \item obfuscation of recipient information for respective external parties (\Cref{e:req:mig,e:req:set}),
    \item delivery of the associated software as a library for embedding into existing applications (\Cref{e:req:dec}), and
    \item optimization of the encryption and decryption procedure to take less than \qty{100}{\milli\second} on modern hardware for common confidential data types (\Cref{e:req:pfm}), while also
    \item be flexible enough to support different cipher suites for future extensions (\Cref{e:req:enc,e:req:key,e:req:pfm}).
\end{itemize}

\begin{figure}[!b]
    \centering%
    \vspace*{-3mm}%
    \begin{tikzpicture}
        \tikzset{
            ecfpart base/.style n args={2}{draw=black, rectangle, minimum width=#2, text width=#2, minimum height=#1, align=center, inner sep=0pt, outer sep=0pt},
            ecfpart/.style={ecfpart base={#1}{5.25cm}, text depth=#1 - 0.55cm}
        }
        \node (ecfbox) at (0, 0) [fill=ecfyellow, ecfpart base={12cm}{5.75cm}, rounded corners=1mm, anchor=north, text depth=12cm - 0.75cm] {\textbf{\acl{ECF}}};
        \node (pubhead) [below=0.75cm of ecfbox.north] [ecfpart=3.5cm, fill=ecfviolet] {Public information about the file and the recipients};
        \node (pribody) [below=0cm of pubhead] [ecfpart=6.25cm, fill=ecfblue] {Private information about the file and the recipients};
        \node (pubfoot) [below=0cm of pribody] [ecfpart=1.25cm, text depth=0, fill=ecfviolet] {Public footer};

        \node (pubhead-recspec) [above left=0cm of pubhead.south east] [ecfpart base={1.5cm}{3.25cm}] {Recipient-specific information for file decryption};
        \node (pubhead-recspec-m) [left=0cm of pubhead-recspec] [ecfpart base={0.75cm}{1.5cm}, rotate=90, anchor=south] {$m$ times};
        \node (pubhead-geninfo) [above left=0cm of pubhead-recspec.north east] [ecfpart base={1cm}{4cm}] {General information for file decrpytion};

        \node (pribody-footer) [above left=0cm of pribody.south east] [ecfpart base={0.75cm}{4cm}] {Private body footer};
        \node (pribody-secdata) [above left=0cm of pribody-footer.north east] [ecfpart base={2cm}{4cm}] {Confidential data\\(passwords, certificate keys, credentials, \dots)};
        \node (pribody-recspec) [above left=0cm of pribody-secdata.north east] [ecfpart base={1.5cm}{3.25cm}] {Information about the recipients for file re-encryption};
        \node (pribody-recspec-n) [left=0cm of pribody-recspec] [ecfpart base={0.75cm}{1.5cm}, rotate=90, anchor=south] {$n$ times};
        \node (pribody-geninfo) [above left=0cm of pribody-recspec.north east] [ecfpart base={1cm}{4cm}] {General information about the file content};

        \pic [ultra thick, line cap=round] at ($(pubhead.west) + (3mm, -6mm)$) {lock=open};
        \pic [ultra thick, line cap=round] at ($(pribody.west) + (3mm, -6mm)$) {lock=closed};
        \pic [ultra thick, line cap=round] at ($(pubfoot.west) + (3mm, -4mm)$) {lock=open};
    \end{tikzpicture}
    \caption{General structure of an \acl{ECF}, cf.~\autocite{Bauer2023}.}\label{fig:general-structure}%
    \vspace*{-1mm}%
\end{figure}

\subsection{Structure, Data Types, Storage Format, and Notation}\label{sec:requirements:structure}
Each \ecf\ comprises a public header, a private body, and a public footer. The general structure of the \ecf\ format is depicted in \Cref{fig:general-structure}. The public fields are unencrypted and therefore readable by anyone. The private body, however, is fully encrypted using a symmetric encryption scheme and can only be decrypted by the recipients of the \ecf. Therefore, external parties do not get access to the confidential data stored within the body of the \ecf. We describe the public and the private components of an \ecf\ in more detail in \Cref{sec:requirements:public,sec:requirements:private}, respectively.

Throughout the following descriptions of the public and private fields of an \ecf, we use this notation to denote the data types of each field:

\begin{itemize}
    \item \dt{u32le}    \tabto{1.5cm} Unsigned \qty{32}{\bit} Integer, Little Endian
    \item \dt{u8}       \tabto{1.5cm} Unsigned \qty{8}{\bit} Integer (byte)
    \item \dt{s}        \tabto{1.5cm} String: length in bytes as \dt{u32le}, then\\
                        \tabto{1.5cm} UTF-8 bytes without byte order mark (BOM)
    \item \dt{x [n]}    \tabto{1.5cm} Array of type \dt{x} with $n$ entries, e.g.,\\
          \dt{u8 [16]}  \tabto{1.5cm} denotes an array of $16$ bytes.
\end{itemize}

All variables referenced in this subsection are listed in \Cref{tab:variables} in \Cref{sec:a:variables}. Our solution employs four types of cryptographic algorithms to achieve the aforementioned goals:

\begin{itemize}
    \item Diffie-Hellman-like Key Agreement/Exchange Algorithm
    \item Matching Signature Scheme (cf.~\Cref{sec:a:general-kex-sig})
    \item Symmetric Encryption Scheme
    \item Suitable Hash Function
\end{itemize}

Each set of these algorithms is called a cipher suite. In our \ac{PoC} implementation, we define four cipher suites listed below. We designed the \ecf\ format to be flexible with regards to the used cipher suite and in order to determine the cipher suite of a specific \ecf, we assigned a unique identifier to each. In comparison to our previous work in \autocite{Bauer2023}, \Cref{e:cip:x25519-ed25519-aegis256-sha256,e:cip:x25519-ed25519-aegis256-sha512} were added to the \ac{PoC} implementation with little effort, proving the flexibility of the \ecf\ format. The four cipher suites are:

\begin{enumerate}[I)]
\setlength{\itemsep}{5pt} 
    \item\label[ciphersuite]{e:cip:x25519-ed25519-aes256gcm-sha256}
          Key Exchange:         \tabto{3.6cm} X25519~\autocite{Bernstein2006}\\
          Signature:            \tabto{3.6cm} Ed25519~\autocite{Bernstein2012}\\
          Symmetric Encryption: \tabto{3.6cm} AES-256-GCM~\autocite{McGrew2005}\autocite{Dworkin2007}\\
          Hash Function:        \tabto{3.6cm} SHA-256~\autocite{Dang2015}\\
          Identifier:           \tabto{3.6cm} \hex{01010101}
    \item\label[ciphersuite]{e:cip:x25519-ed25519-aes256gcm-sha512}
          Key Exchange:         \tabto{3.6cm} X25519~\autocite{Bernstein2006}\\
          Signature:            \tabto{3.6cm} Ed25519~\autocite{Bernstein2012}\\
          Symmetric Encryption: \tabto{3.6cm} AES-256-GCM~\autocite{McGrew2005}\autocite{Dworkin2007}\\
          Hash Function:        \tabto{3.6cm} SHA-512~\autocite{Dang2015}\\
          Identifier:           \tabto{3.6cm} \hex{01010102}
    \item\label[ciphersuite]{e:cip:x25519-ed25519-aegis256-sha256}
          Key Exchange:         \tabto{3.6cm} X25519~\autocite{Bernstein2006}\\
          Signature:            \tabto{3.6cm} Ed25519~\autocite{Bernstein2012}\\
          Symmetric Encryption: \tabto{3.6cm} AEGIS-256~\autocite{Denis2023}\\
          Hash Function:        \tabto{3.6cm} SHA-256~\autocite{Dang2015}\\
          Identifier:           \tabto{3.6cm} \hex{01010201}
    \item\label[ciphersuite]{e:cip:x25519-ed25519-aegis256-sha512}
          Key Exchange:         \tabto{3.6cm} X25519~\autocite{Bernstein2006}\\
          Signature:            \tabto{3.6cm} Ed25519~\autocite{Bernstein2012}\\
          Symmetric Encryption: \tabto{3.6cm} AEGIS-256~\autocite{Denis2023}\\
          Hash Function:        \tabto{3.6cm} SHA-512~\autocite{Dang2015}\\
          Identifier:           \tabto{3.6cm} \hex{01010202}
\end{enumerate}

The provided \ac{PoC} contains a command line tool, which uses \Cref{e:cip:x25519-ed25519-aes256gcm-sha512} by default. When creating an \ecf\ using the command line tool, the user can override the cipher suite to use. There is currently no built-in option to change the cipher suite afterwards.
This is because some combinations of cipher suites -- in particular the keys for key exchange and signing -- may be incompatible. This problem does not affect the implemented ones, but may arise due to future extensions.

\subsection{Unencrypted, Public Header and Footer Fields}\label{sec:requirements:public}
All authorized recipients must be able to decrypt the file and, thus, must be able to acquire the symmetric encryption key used to encrypt the private body. Simultaneously, in order to achieve the goal of minimal public information about the recipients, the public header fields contain as little identifying information as possible. Our solution allows us to separate the information needed for decrypting and the information needed for encrypting. Whereas the former must be made public in order for the whole scheme to work, the latter can be included in the encrypted body and therefore be protected against unauthorized access.

The public header of an \ecf\ comprises a general, recipient-independent part followed by $m$ identically constructed recipient-specific blocks. \Cref{fig:public-header} shows all public header fields in their respective order. The recipient-independent fields contain general information about the \ecf:

\begin{fielditemize}
    \item \fdt{Container Version}{u32le}\\\ecf\ format version; intended for future extensions;\\currently set to \hex{00010000} for \enquote{Version 1.0}
    \item \fdt{Cipher Suite}{u32le}\\Identifier for chosen cipher suite;\\cf. enumeration in \Cref{sec:requirements:structure}
    \item \fdt{Public Length}{u32le}\\Length $h$ of the public header in bytes
    \item \fdt{Private Length}{u32le}\\Length $b$ of the encrypted private body in bytes
    \item \fdt{Recipient Count}{u32le}\\Number of recipients $m$ in the public header
    \item \fdt{Salt}{u8 [16]}\\Salt value; cf. \Cref{sec:ops:create,sec:ops:use}
    \item \fdt{Symmetric Nonce}{u8 [c]}\\Symmetric nonce value; cf. \Cref{sec:ops:create,sec:ops:use};\\Length dependent on chosen cipher suite, usually $c=12$
\end{fielditemize}

\begin{figure}[htb]
    \centering%
    \vspace*{-3mm}%
    \begin{tikzpicture}
        \node (cv) at (0, 0) [u32le] {\field{Container Version}};
        \node (cs) [right=0cm of cv] [u32le] {\field{Cipher Suite}};
        \node (pu) [right=0cm of cs] [u32le] {\field{Public Length} $h$};
        \node (pr) [right=0cm of pu] [u32le] {\field{Private Length} $b$};
        \node (rc) [below=0cm of cv] [u32le] {\field{Recipient Count} $m$};
        \node (s1) [right=0cm of rc] [u8a={12}] {\dt{u8 [16]} \field{Salt}~\dots};
        \node (s2) [below=0cm of rc] [u8a={4}] {\dots~\field{Salt}};
        \node (sn) [right=0cm of s2] [u8a={12}] {\dt{u8 [c]} \field{Symmetric Nonce}};
        \node (b1) [below right=0cm of s2.south west] [u8a={16}] {\dt{dblock [m]} \field{Recipient-Specific Blocks}~\dots};
        \node (b2) [below=0cm of b1] [u8a={16}, half height] {\dots};
        \node (b3) [below=0cm of b2] [u8a={16}, half height] {\dots};
        \node (b4) [below=0cm of b3] [u8a={16}, half height] {\dots};
        \node (b5) [below=0cm of b4] [u8a={16}] {\dots~\field{Recipient-Specific Blocks}};

        \node [above right=0cm and \bitshoffset cm of cv.north west] {$0$};
        \node [above=0cm of cv.north east] {$4$};
        \node [above=0cm of cs.north east] {$8$};
        \node [above=0cm of pu.north east] {$12$};
        \node [above left=0cm and \bitshoffset cm of pr.north east] {$16$};
        \node [below left=0cm and \bitshoffset cm of b5.south east] {$h$};
    \end{tikzpicture}
    \vspace*{-3mm}%
    \caption{Public header fields of an \acl{ECF}.}\label{fig:public-header}%
    \vspace*{-1mm}%
\end{figure}

All $4$~byte-long fields are of type \dt{u32le}. While the length of \field{Salt} is fixed to $16$~bytes, the length of \field{Symmetric Nonce} is dependent on the chosen cipher suite. However, since all implemented cipher suites use an \ac{AEAD} symmetric encryption scheme, we follow the recommendations in RFC~5116~\autocite{McGrew2008} and set it to $c=12$~bytes as shown in \Cref{fig:public-header} for the \ac{GCM}~\autocite{McGrew2005} of \ac{AES}. The AEGIS specification~\autocite{Denis2023}, however, requires $c=y$.

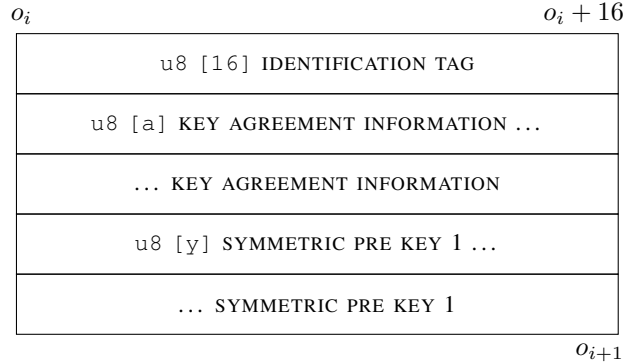
\begin{figure}[htb]
    \centering%
    \vspace*{-3mm}%
    \begin{tikzpicture}
        \node (id) at (0, 0) [u8a={16}] {\dt{u8 [16]} \field{Identification Tag}};
        \node (x1) [below=0cm of id] [u8a={16}] {\dt{u8 [a]} \field{Key Agreement Information}~\dots};
        \node (x2) [below=0cm of x1] [u8a={16}] {\dots~\field{Key Agreement Information}};
        \node (k1) [below=0cm of x2] [u8a={16}] {\dt{u8 [y]} \field{Symmetric Pre Key 1}~\dots};
        \node (k2) [below=0cm of k1] [u8a={16}] {\dots~\field{Symmetric Pre Key 1}};

        \node [above right=0cm and \bitshoffset cm of id.north west] {$o_i$};
        \node [above left=0cm and \bitshoffset cm of id.north east] {$o_i + 16$};
        \node [below left=0cm and \bitshoffset cm of k2.south east] {$o_{i+1}$};
    \end{tikzpicture}
    \vspace*{-2mm}%
    \caption{Recipient-specific block structure within the public header of an \acl{ECF}.}\label{fig:public-recipient-block}%
    \vspace*{-2mm}%
\end{figure}

There are exactly $m$ recipient-specific decryption blocks or \dt{dblock}s following the recipient-independent part. The structure and length of these blocks is also highly dependent on the cipher suite. \Cref{fig:public-recipient-block} shows the general, three-part structure of these blocks. During the decryption procedure, recipients have to find their respective block by matching a unique identification number. This is why \field{Identification Tag} is the first entry in each block and has a fixed length of $16$~bytes. It allows for up to $(2^8)^{16} = 2^{128}$, i.e., practically unlimited number of recipients while saving space compared to a full-length hash value. For illustration purposes, we decided to use the array lengths of the default \Cref{e:cip:x25519-ed25519-aes256gcm-sha512}. In this case, \field{Key Agreement Information} contains an ephemeral X25519 public key of length $a=32$~bytes and \field{Symmetric Pre Key 1} has, because of AES-256, a length of $y=32$~bytes.

In \Cref{fig:public-recipient-block}, we use $o_i$ to denote the offset of a recipient-specific block $0 \le i < m$ in bytes measured from the start of the file. It can be calculated as $o_i = i \cdot (16 + a + y) + o_0$. Since the blocks follow the recipient-independent part, $o_0 = 36 + c$. Furthermore, $o_m = h$ must hold in order for an \ecf\ to be considered valid.

As shown in \Cref{fig:general-structure}, an \ecf\ also contains a public footer. It consists of a single field: \fdt{File Hash}{u8 [d]}. Its length is dependent on the hash function of the chosen cipher suite. For example, $d=64$~bytes applies to SHA-512. The field \field{File Hash} serves primarily to check the integrity of the file, e.g., after a copy or transfer action, and, thus, helps to detect accidental alterations. It does not increase the resilience against a malicious attacker as it can be recomputed easily after a deliberate alteration of other parts of the file.

The encryption procedure fills or generates all public header and footer fields and the decryption procedure uses them. Both operations are described in detail in \Cref{sec:ops}.

\subsection{Encrypted, Private Body Fields}\label{sec:requirements:private}
After retrieving the symmetric encryption key using the public header information, an authorized recipient is able to decrypt the private body of the \ecf. External parties, who are not recipients of that \ecf, are unable to obtain the encryption key and, thus, unable to access any information stored within the body of that \ecf. The body of an \ecf\ consists of four parts as shown in \Cref{fig:general-structure}. The fields within the \ecf\ body, which are depicted in \Cref{fig:private-body}, are composed as follows:

\begin{fielditemize}
    \item \fdt{Content Type}{u32le}\\Describes the type of the confidential data
    \item \fdt{Public Header Hash}{u8 [d]}\\Hash value of the public header
    \item \fdt{Recipient Count}{u32le}\\Number of true recipients $n \le m$
    \item \fdt{Recipient Information Blocks}{rblock [n]}\\Information about the recipients of that \ecf
    \item \fdt{Content Length}{u32le}\\Length $q$ of the confidential data in bytes
    \item \fdt{Content}{u8 [q]}\\Confidential data
    \item \fdt{Private Hash}{u8 [d]}\\Hash value of the private body so far
\end{fielditemize}

\begin{figure}[!b]
    \centering%
    \vspace*{-4mm}%
    \begin{tikzpicture}
        \node (ct) at (0, 0) [u32le] {\field{Content Type}};
        \node (h1) [right=0cm of ct] [u8a={12}] {\dt{u8 [d]} \field{Public Header Hash}~\dots};
        \node (h2) [below right=0cm of ct.south west] [u8a={16}, half height] {\dots};
        \node (h3) [below right=0cm of h2.south west] [u8a={4}] {\dots~\field{PH Hash}};
        \node (rc) [right=0cm of h3] [u32le] {\field{Recipient Count} $n$};
        \node (b1) [right=0cm of rc] [u8a={8}] {\dt{rblock [n]}\\\field{Recipient Information}~\dots};
        \node (b2) [below right=0cm of h3.south west] [u8a={16}, half height] {\dots};
        \node (b3) [below=0cm of b2] [u8a={16}, half height] {\dots};
        \node (b4) [below=0cm of b3] [u8a={16}, half height] {\dots};
        \node (b5) [below=0cm of b4] [u8a={16}] {\dots~\field{Recipient Information Blocks}};
        \node (cl) [below right=0cm of b5.south west] [u32le] {\field{Content Length} $q$};
        \node (c1) [right=0cm of cl] [u8a={12}] {\dt{u8 [q]} \field{Content}~\dots};
        \node (c2) [below right=0cm of cl.south west] [u8a={16}, half height] {\dots};
        \node (c3) [below=0cm of c2] [u8a={16}, half height] {\dots};
        \node (c4) [below=0cm of c3] [u8a={16}, half height] {\dots};
        \node (c5) [below=0cm of c4] [u8a={16}] {\dots~\field{Content}};
        \node (g1) [below=0cm of c5] [u8a={16}] {\dt{u8 [d]} \field{Private Body Hash}~\dots};
        \node (g2) [below=0cm of g1] [u8a={16}, half height] {\dots};
        \node (g3) [below=0cm of g2] [u8a={16}] {\dots~\field{Private Body Hash}};

        \node [above right=0cm and \bitshoffset cm of ct.north west] {$0$};
        \node [above left=0cm and \bitshoffset cm of h1.north east] {$16$};
        \node [below left=0cm and \bitshoffset cm of g3.south east] {$b'$};
    \end{tikzpicture}
    \caption{Private body fields of an \acl{ECF}.}\label{fig:private-body}%
    \vspace*{-1mm}%
\end{figure}

The field \field{Content Type} characterizes the type of confidential data stored within \field{Content}. Our \ac{PoC} implementation provides a built-in type of content, \ac{BLOB}, using the identifier \hex{00000001}. Since our implementation is flexible and extensible, future applications can and should define and handle their own types of content.

Similar to the public footer described in \Cref{sec:requirements:public}, the lengths of the two fields \field{Public Header Hash} and \field{Private Body Hash} is dependent on the chosen cipher suite. For \Cref{e:cip:x25519-ed25519-aes256gcm-sha512}, $d=64$~bytes. Both fields aim to ensure integrity of the \ecf\ and to detect unauthorized or unintended modifications of both the public header and the private body. For example, if one of the $m$ public, recipient-specific blocks is altered preventing the corresponding recipient from obtaining the symmetric encryption key, all other recipients can detect this alteration. We opted to use fields for the hash values in order to allow for non-\ac{AEAD}-based symmetric encryption schemes. This is also why we do not make use of the \enquote{associated data} option during encryption and decryption (cf.~\Cref{sec:ops}).

Next, the private body contains information about the recipients in variable-sized recipient information blocks or \dt{rblock}s. There are exactly $n$ of these blocks present as denoted in \Cref{fig:private-body} and they contain the necessary information for file re-encryption. Allowing for $m \ge n$ is a design choice we made to hide the true number of recipients $n$ from external parties, who can only read the public header field \field{Recipient Count} and, thus, only obtain $m$. Furthermore, by storing the recipient-identifying information within the encrypted part of an \ecf, we ensure that this information is only accessible by the recipients themselves and hidden from external parties, thus, pursuing the goal of minimal publicly available information about the recipients.

\begin{figure}[!b]
    \centering%
    \begin{tikzpicture}
        \node (pk1) at (0, 0) [u8a={16}] {\dt{u8 [u]} \field{Public Key}~\dots};
        \node (pk2) [below=0cm of pk1] [u8a={16}] {\dots~\field{Public Key}};
        \node (name1) [below=0cm of pk2] [u8a={16}] {\dt{s} \field{Name}~\dots};
        \draw [dashed] ($(name1.north west) + (\uintwidth cm, 0)$) -- +(0, -\blockheight cm);
        \node (name2) [below=0cm of name1] [u8a={16}] {\dots~\field{Name}};
        \node (sig1) [below=0cm of name2] [u8a={16}] {\dt{u8 [g]} \field{Name Signature}~\dots};
        \node (sig2) [below=0cm of sig1] [u8a={16}, half height] {\dots};
        \node (sig3) [below=0cm of sig2] [u8a={16}] {\dots~\field{Name Signature}};

        \node [above right=0cm and \bitshoffset cm of pk1.north west] {$0$};
        \node [above left=0cm and \bitshoffset cm of pk1.north east] {$16$};
        \node [below left=0cm and \bitshoffset cm of sig3.south east] {$u + \left|\field{Name}\right| + g$};
    \end{tikzpicture}
    \caption{Recipient information block structure within the private body of an \acl{ECF}.}\label{fig:private-recipient-block}%
    \vspace*{-1mm}%
\end{figure}
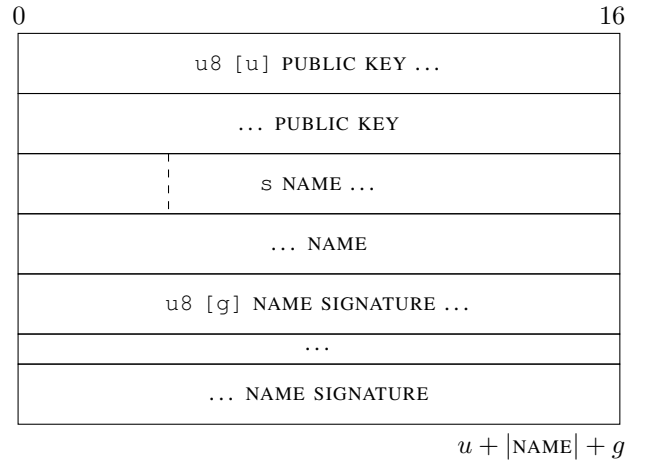

Each \dt{rblock} contains three fields as shown in \Cref{fig:private-recipient-block}:

\begin{fielditemize}
    \item \fdt{Public Key}{u8 [u]}\\Public key of the recipient;\\Length dependent on chosen cipher suite, e.g., $u=32$
    \item \fdt{Name}{s}\\Self-chosen name of the recipient, usually email address or another organization-specific identifier
    \item \fdt{Name Signature}{u8 [g]}\\Signature over \field{Name};\\Length dependent on chosen cipher suite, e.g., $g=64$
\end{fielditemize}

Since these blocks do not have a fixed size due to the variable-length self-chosen \field{Name}, the offset measured from the start of the private body must be calculated by iterating through all previous blocks. The signature over the self-chosen name serves to assure that the person owning the associated private key has chosen the name and that no changes have been made to the name afterwards. The name is for information purposes only, e.g., when displaying the recipients or removing an existing recipient by name.

The main motivation for designing the \ecf\ format was to securely store confidential data in a multi-recipient setting. The secrets, which may have any format the users want them to, are stored within the field \field{Content}. Its theoretical maximum size $\max(q)$ in bytes is ultimately dependent on the number of true recipients $n$ and the lengths of their self-chosen names as well as the chosen cipher suite. This is because the length of the encrypted private is limited to \mbox{$\max(b) = 2^{32} - 1 \mathop{}\textnormal{bytes} \approx \qty{4}{\gibi\byte}$}. We can calculate the theoretical maximum size of the confidential data as follows:
\begin{align*}
    \max(q) &= \max(b) - \bar{q} - \left(b - b'\right)\\
    \bar{q} &= b' - q\\[-1mm]
            &= 4 + d + 4 + \sum_{i=1}^{n} \left(u + \left|\field{Name}_i\right| + g\right) + 4 + d\\[-1mm]
            &= 12 + 2d + n\left(u + g\right) + \sum_{i=1}^{n} \left|\field{Name}_i\right|
\end{align*}

The total length of all private fields is denoted by $b' \le b$, whereas $b$ denotes the length of the encrypted private body. The public header field \field{Private Length} holds the value $b$, which may be greater than $b'$ depending on the used symmetric encryption scheme. This is definitely true for \ac{AEAD}-based algorithms. Therefore, $b - b' \ge 0$ describes the number of bytes the encrypted private body exceeds the unencrypted one. The helper variable $\bar{q} = b' - q > 0$ denotes the length of all private body fields except \field{Content}.

We can show that $\bar{q} + (b - b')$ is negligible compared to $\max(b)$, and therefore the amount of data that can be stored within an \ecf\ is sufficient for passwords, certificate keys, and similar confidential data. We assume $n=1000$ recipients with an average name length of $\left|\field{Name}_i\right| = 500$~bytes and using the default \Cref{e:cip:x25519-ed25519-aes256gcm-sha512}:
\begin{align*}
    \bar{q} + (b - b') &= 12 + 2\cdot 64 + 1000 \cdot \left(32 + 64 + 500\right) + 16\\
                       &= 596157 \approx 2^{19} \ll 2^{32} - 1 = \max(b)\\
    \implies \max(q) &\approx 2^{32} - 2^{19} \approx 2^{32}
\end{align*}

In practice, both $n$ and the average name length will be significantly less, supporting our argument even more. Our strong assumption is that, in practice, the number of recipients of a single \ecf\ does typically not exceed $20$. Furthermore, using email addresses as recipient identification in the field \field{Name} as we suggested earlier yields far shorter names than in the assumption above.

\section{\ecf\ Operations}\label{sec:ops}
This section describes the basic operations on an \ecf. We follow roughly the workflow presented in \Cref{sec:app} and describe the encryption procedure in \Cref{sec:ops:create}, the decryption procedure in \Cref{sec:ops:use}, and the modification of the recipient set and confidential data in \Cref{sec:ops:recipients,sec:ops:content}, respectively.

\textit{Nomenclature.} We use a common notation for the following subsections. Since the \ecf\ format is flexible with regards to the used cipher suite, we opted to use variable and function names that are not tied to a specific algorithm. Due to readability concerns, we moved the notation to \Cref{sec:a:variables}, and specifically to \Cref{tab:component-variables,,tab:functions}.

\subsection{Creating and Encrypting an \ecf}\label{sec:ops:create}
As described in \Cref{sec:app:create}, Alice first needs to generate a key pair consisting of a private and a public key. All cipher suites in our \ac{PoC} implementation allow to convert an Ed25519 key pair to a X25519 key pair as described in~\autocite{CSE2019}\autocite{Sodium2022ed2c}. Therefore, we generate the key pair for signing and then convert it to a key pair for key exchange:
\begin{align*}
    \left(\privatekey{A}{S}, \publickey{A}{S}\right) &\leftarrow \gen^\textnormal{S} \\
    \left(\privatekey{A}{X}, \publickey{A}{X}\right) &= \left(\convertSX\left(\privatekey{A}{S}\right), \convertSX\left(\publickey{A}{S}\right)\right)
\end{align*}
Two components are essential to an \ecf: the confidential data $\Omega \in \left\{0, 1\right\}^{8q}$ and the set of recipients $\Psi = \left\{\psi_0, \dots, \psi_{n-1}\right\}$. When creating a new \ecf, both components are empty. As a next step, Alice adds herself to the recipient set and, therefore, must generate a recipient entry $\psi_a = \publickey{A}{S} \Vert \mfield{Name}_a \Vert t_a$. She may choose a name or use, for example, her email address as an identifier. Then she signs it:
\[t_a = \sign\left(\privatekey{A}{S}, \utf\left(\mfield{Name}_a\right)\right)\]

Alice now adds herself to the recipient set, i.e., $\Psi = \left\{\psi_a\right\}$. Following the basic workflow described earlier, Alice takes the byte representation of the confidential data in any format she likes and assigns it to $\Omega$. As a final step, Alice encrypts the two components, thus, obtaining a valid \ecf:
\[ \mathcal{E} \leftarrow \encrypt^\textnormal{ECF}\left(\Psi, \Omega\right)\]

The encryption procedure takes two inputs, the set of recipients~$\Psi$ and the confidential data $\Omega$. It is important to note that parts of the cipher suite are selected implicitly by the construction of $\psi_a$. In the following, we assume that all cipher suite algorithms and their respective properties, e.g., the symmetric key length $y$ and the nonce length $c$, are well defined according to the cipher suite Alice chose.

Additionally, the encryption procedure does not require any private key directly. However, the generation of $\psi_a$ did require Alice's private key for signing, which is why our \ac{PoC} implementation requests access to it. Yet, e.g., a pre-created recipient set~$\Psi$, allows anyone to create an \ecf\ without first generating a key pair. This operation, however, is not supported by our \ac{PoC} implementation as the creator would not be a recipient and, thus, be unable to access the content of that~\ecf.

The encryption procedure is defined as follows:
\Crefname{enumi}{Step}{Steps}
\crefname{enumi}{step}{steps}
\begin{enumerate}[(1)]
    \setlength{\itemsep}{5pt} 
    \item\label{e:enc:key-nonce} Generate a symmetric key, a nonce, and a salt\\$\symmetrickey{final} \leftarrow \gen^\textnormal{SYM}, \alpha \leftarrow \random(c), \theta \leftarrow \random(16)$.
    \item\label{e:enc:choose-m} Choose a random integer $m \ge n = \left|\Psi\right|$.
    \item\label{e:enc:gen-rdi} For all $\psi_i \in \Psi, 0 \le i < n$ generate $r_i$: ($R = \emptyset$ initially)
    \begin{enumerate}[(a)]
        \item\label{e:enc:gen-rdi:id-tag} Compute \field{Identification Tag}\\$\textnormal{id\_tag} = \hash\left(\publickey{$\psi_i$}{S} \Vert \theta\right)\trunc{0}{15}$.
        \item\label{e:enc:gen-rdi:gen-eph} Generate an ephemeral key pair for key exchange\\$\left(\privatekey{e}{X}, \publickey{e}{X}\right) \leftarrow \gen^\textnormal{X}$.
        \item\label{e:enc:gen-rdi:cvt} Convert the recipient's public key for signing to a public key for key exchange $\publickey{$\psi_i$}{X} = \convertSX\left(\publickey{$\psi_i$}{S}\right)$.
        \item\label{e:enc:gen-rdi:kex} Execute the key exchange algorithm:\\$\sharedsecret = \keyexchange\left(\privatekey{e}{X}, \publickey{$\psi_i$}{X}\right)$.
        \item\label{e:enc:gen-rdi:pre2} Calculate $\symmetrickey{pre2} = \hash\left(\sharedsecret \Vert \publickey{$\psi_i$}{X} \Vert \publickey{e}{X}\right)\trunc{0}{y-1}$.\\Using this hash function construction instead of the shared secret directly is recommended in~\autocite{Sodium2022psm}.
        \item\label{e:enc:gen-rdi:pre1} Calculate $\symmetrickey{pre1} = \symmetrickey{final} \xor \symmetrickey{pre2}$.
        \item\label{e:enc:gen-rdi:fin} Add $r_i = \textnormal{id\_tag} \Vert \publickey{e}{X} \Vert \symmetrickey{pre1}$ to $R$.
    \end{enumerate}
    \item\label{e:enc:gen-fake} Generate $m - n$ deception entries as described in \Cref{sec:ops:deception}, add them to $R$ and shuffle or sort the elements $r_i \in R, 0 \le i < m$ by $\textnormal{id\_tag}$.
    \item\label{e:enc:write-public} Write all public header fields into a temporary buffer $\mathcal{H}^*$, except for \field{Private Length} as it is not known yet. Set its value $b$ to \hex{ECFFC0DE} (\ecf\ format \enquote{code}) and compute the hash value of the public header:
    \begin{align*}
        \mathcal{H}^* &= \mfield{Container Version} \Vert \mfield{Cipher Suite} \Vert h \Vert\\&\qquad\hex{ECFFCODE} \Vert m \Vert \theta \Vert \alpha \Vert r_1 \Vert r_2 \Vert \dots \Vert r_m \\
        \beta_\textnormal{public} &= \hash\left(\mathcal{H}^*\right).
    \end{align*}
    \item\label{e:enc:write-private} Write all private body fields unencrypted into a temporary buffer $\mathcal{B}'$ with $q = \left|\Omega\right|$ in bytes and compute the hash value of the private body:
    \begin{align*}
        \psi_i &= \publickey{$\psi_i$}{S} \Vert \mfield{Name}_i \Vert t_i \\
        \mathcal{B}' &= \mfield{Content Type} \Vert \beta_\textnormal{public} \Vert n \Vert \\ &\qquad\quad \psi_1 \Vert \psi_2 \Vert \dots \Vert \psi_n \Vert q \Vert \Omega \\
        \beta_\textnormal{private} &= \hash\left(\mathcal{B}'\right).
    \end{align*}
    \item\label{e:enc:encrypt} Append $\beta_\textnormal{private}$ to $\mathcal{B}'$ and then encrypt the buffer:
    \begin{align*}
        \mathcal{B} \leftarrow \encrypt^\textnormal{SYM}\left(\symmetrickey{final}, \alpha, \mathcal{B}' \Vert \beta_\textnormal{private}\right).
    \end{align*}
    \item\label{e:enc:fin} Update the previously written field \field{Private Length} in $\mathcal{H}^*$ with its actual value $b = \left|\mathcal{B}\right|$ in bytes to obtain the final header $\mathcal{H}$ and compute the hash value over everything:
    \begin{align*}
        \mathcal{H} &= \mfield{Container version} \Vert \mfield{Cipher Suite} \Vert h \Vert \\ &\qquad\qquad\qquad\quad b \Vert m \Vert \theta \Vert \alpha \Vert r_1 \Vert r_2 \Vert \dots \Vert r_m \\
        \beta_\textnormal{all} &= \hash\left(\mathcal{H} \Vert \mathcal{B}\right).
    \end{align*}
    \item\label{e:enc:out} The output $\mathcal{E}$ of the \ecf\ encryption procedure is
    \begin{align*}
        \mathcal{E} = \mathcal{H} \Vert \mathcal{B} \Vert \beta_\textnormal{all}.
    \end{align*}
\end{enumerate}

\subsection{Decrypting an \ecf}\label{sec:ops:use}
Only recipients of an \ecf\ are able to decrypt it. In particular, having access to the private key of a recipient's key pair implies access to the encrypted content stored in the private body of an \ecf. Alice has to use her private key for signing in order to decrypt the \ecf\ and obtain the recipient set $\Psi$ and the confidential data $\Omega$:
\[ \left(\Psi, \Omega\right) = \decrypt^\textnormal{ECF}\left(\privatekey{A}{S}, \mathcal{E}\right) \]

The decryption procedure is defined as follows:
\begin{enumerate}[(1)]
    \setlength{\itemsep}{4pt} 
    \item\label{e:dec:load} Load the public header fields and determine the container version as well as the cipher suite. Now $h, b, m, \theta$, and~$\alpha$ have values. Let $\mathcal{H}$ be the public header, $\mathcal{B}$ be the encrypted private body, and $\beta_\textnormal{all}$ the public footer.
    \item\label{e:dec:verify-filehash} Check file integrity. If $\verify^\textnormal{H}\left(\mathcal{H} \Vert \mathcal{B}, \beta_\textnormal{all}\right) \overset{?}{=} 0$, \textit{Exit}.
    \item\label{e:dec:cvt} Compute all necessary keys:\\
    $\privatekey{A}{X} = \convertSX\left(\privatekey{A}{S}\right)$,\\$\publickey{A}{S} = \convertskpk\left(\privatekey{A}{S}\right)$, $\publickey{A}{X} = \convertskpk\left(\privatekey{A}{X}\right)$.
    \item\label{e:dec:id-tag} Compute $\textnormal{id\_tag} = \hash\left(\publickey{A}{S} \Vert \theta\right)\trunc{0}{15}$.
    \item\label{e:dec:search} Search recipient-specific block $r_a = \textnormal{id\_tag} \Vert \publickey{e}{X} \Vert \symmetrickey{pre1}$ with matching $\textnormal{id\_tag}$.\\
    \hspace*{4mm}If not found: Alice is not a recipient of this \ecf. \textit{Exit}.
    \item\label{e:dec:kex} Execute the key exchange algorithm:\\$\sharedsecret = \keyexchange\left(\privatekey{A}{X}, \publickey{e}{X}\right)$.
    \item\label{e:dec:pre2} Calculate $\symmetrickey{pre2} = \hash\left(\sharedsecret \Vert \publickey{A}{X} \Vert \publickey{e}{X} \right)\trunc{0}{y-1}$.
    \item\label{e:dec:pre1} Calculate $\symmetrickey{final} = \symmetrickey{pre1} \xor \symmetrickey{pre2}$.
    \item\label{e:dec:decrypt} Decrypt the private body $\mathcal{B}$ using the computed symmetric key and the nonce $\alpha$ into a buffer $\mathcal{B}^*$:\vspace*{-1.5mm}
    \[ \mathcal{B}^* = \decrypt^\textnormal{SYM}\left(\symmetrickey{final}, \alpha, \mathcal{B}\right). \]
    \item\label{e:dec:load-private} Deconstruct $\mathcal{B}^* = \mathcal{B}' \Vert \beta_\textnormal{private}$ into the private body fields.
    \item\label{e:dec:verify-pubhash} Compute the hash over the public header fields with $b$ set to \hex{ECFFCODE} and verify it:\vspace*{-1.5mm}
    \begin{align*}
        \mathcal{H}^* &= \mfield{Container Version} \Vert \mfield{Cipher Suite} \Vert h \Vert\\&\qquad\hex{ECFFCODE} \Vert m \Vert \theta \Vert \alpha \Vert r_1 \Vert r_2 \Vert \dots \Vert r_m\\
        \textnormal{If~}&\verify^\textnormal{H}\left(\mathcal{H}^*, \beta_\textnormal{public}\right) \overset{?}{=} 0 \textnormal{, \textit{Exit}.}
    \end{align*}
    \item\label{e:dec:verify-signatures} For each recipient $\psi_i \in \Psi$:
    \begin{enumerate}[(a)]
        \item\label{e:dec:verify-signatures:load} Load and deconstruct $\psi_i$ into $\publickey{$\psi_i$}{S}$, $\mfield{Name}_i$, and $t_i$.
        \item\label{e:dec:verify-signatures:verify} Verify the signature:\vspace*{-1.5mm}
        \[ \textnormal{If~} \verify^\textnormal{S}\left(\publickey{$\psi_i$}{S},  \utf\left(\mfield{Name}_i\right), t_i\right) \overset{?}{=} 0 \textnormal{, \textit{Exit}.} \]
    \end{enumerate}
    \item\label{e:dec:verify-privhash} Compute the hash over the unencrypted private body excluding the field \field{Private Body Hash} and verify it:\vspace*{-1.5mm}
    \begin{align*}
        \mathcal{B}' &= \mathcal{B}^*\trunc{0}{b' - d} \\
                     &= \mfield{Content Type} \Vert \beta_\textnormal{public} \Vert n \\
                     &\qquad\quad \psi_1 \Vert \psi_2 \Vert \dots \Vert \psi_n \Vert q \Vert \Omega \\
        \textnormal{If~}&\verify^\textnormal{H}\left(\mathcal{B}', \beta_\textnormal{private}\right) \overset{?}{=} 0 \textnormal{, \textit{Exit}.}
    \end{align*}
    \item\label{e:dec:out} The outputs of the \ecf\ decryption procedure are the recipient set $\Psi$ and the confidential data $\Omega$.
\end{enumerate}

Performing \Cref{e:dec:verify-filehash,e:dec:verify-pubhash,e:dec:verify-signatures,,e:dec:verify-privhash} is optional but strongly recommended in order to detect unintended and/or malicious modifications to the \ecf. However, \Cref{e:dec:verify-signatures} takes a considerable amount of time and therefore may be omitted in certain cases. For example, a \ac{CI} pipeline usually contains a build stage and a test stage~\autocite{Fowler2006}. We assume that the whole pipeline runs in a trusted environment, e.g., a self-hosted server. Then, it is advisable to check the recipient's signatures as a prerequisite to the build stage, while this step may be skipped during testing in order to speed up the pipeline execution. Nevertheless, the execution times in question are in the range of milliseconds, cf.~\Cref{sec:improvements:perf}.

\subsection{Adding and Removing Recipients}\label{sec:ops:recipients}
Adding recipients to an \ecf\ or removing recipients from an \ecf\ implies modifying the recipient set $\Psi$. We follow the basic workflow described in \Cref{sec:app:add,sec:app:remove}. In order to obtain $\Psi$ (and the confidential data $\Omega$), Alice is required to be a recipient of the \ecf\ $\mathcal{E}$, i.e. $\psi_a \in \Psi$.

In the following, we describe the recipient addition procedure for a single recipient: Bob. Adding more recipients, such as Charlie and the automated deployment job, is performed analogously.

Before Alice can add Bob to $\mathcal{E}$, her coworker has to generate his recipient entry $\psi_b$ and, thus, has to generate key pairs analogous to what Alice did in \Cref{sec:ops:create}:
\begin{align*}
    \left(\privatekey{B}{S}, \publickey{B}{S}\right) &\leftarrow \gen^\textnormal{S} \\
    \left(\privatekey{B}{X}, \publickey{B}{X}\right) &= \left(\convertSX\left(\privatekey{B}{S}\right), \convertSX\left(\publickey{B}{S}\right)\right) \\
    t_b &= \sign\left(\privatekey{B}{S}, \utf\left(\mfield{Name}_b\right)\right) \\
    \psi_b &= \publickey{B}{S} \Vert \mfield{Name}_b \Vert t_b
\end{align*}

Bob now sends $\psi_b$ to Alice. Alice performs the following steps to add him as a recipient to $\mathcal{E}$:
\begin{enumerate}[(1)]
    \setlength{\itemsep}{4pt} 
    \item\label{e:add:verify} Verify signature:\vspace*{-1.5mm}
    \[ \textnormal{If~} \verify^\textnormal{S}\left(\publickey{$\psi_b$}{S}, \utf\left(\mfield{Name}_b\right), t_b\right) \overset{?}{=} 0 \textnormal{, \textit{Exit}.} \]
    \item\label{e:add:dec} Decrypt the \ecf: $\left(\Psi, \Omega\right) = \decrypt^\textnormal{ECF}\left(\privatekey{A}{S}, \mathcal{E}\right)$.
    \item\label{e:add:check-pubk} Check, if Bob is already a recipient by comparing the public keys for signing: If $\exists\,\psi_i \in \Psi: \publickey{$\psi_i$}{S} = \publickey{B}{S}$, \textit{Exit}.
    \item\label{e:add:check-name} Optionally check, if Bob's name is already present in the recipient set: If $\exists\,\psi_i \in \Psi: \mfield{Name}_i = \mfield{Name}_b$, \textit{May exit}.
    \item\label{e:add:add} Add Bob: $\Psi' = \Psi \cup \left\{\psi_b\right\} = \left\{\psi_0, \psi_1, \dots, \psi_{n-1}, \psi_b\right\}$.
    \item\label{e:add:enc} Encrypt the modified \ecf: $\mathcal{E}' \leftarrow \encrypt^\textnormal{ECF}\left(\Psi', \Omega\right)$.
\end{enumerate}
The name of a recipient is for information purposes only, which results in \Cref{e:add:check-name} to be optional. However, in order to avoid (human) confusion about the recipients of an \ecf, it is advisable to enforce the constraint that all recipient names must be unique. Our \ac{PoC} implementation does not allow duplicate names by default but can be configured otherwise.

We now assume that Alice added her coworkers Bob and Charlie as well as the automated deployment job as recipients to the \ecf, thus, following the basic workflow described in \Cref{sec:app:add}.

Next, Alice wants to remove Bob from the recipient set as motivated in \Cref{sec:app:remove}. She has to identify Bob in the recipient set either by his public key or his name. If Alice uses Bob's name to identify him, his name must be unique within the \ecf. Alice performs the following steps to remove Bob from the recipient set:
\begin{enumerate}[(1)]
    \setlength{\itemsep}{4pt} 
    \item\label{e:rem:dec} Decrypt the \ecf: $\left(\Psi, \Omega\right) = \decrypt^\textnormal{ECF}\left(\privatekey{A}{S}, \mathcal{E}\right)$.
    \item\label{e:rem:identify} Find $\psi_b \in \Psi$ based on $\publickey{B}{S}$ and/or $\mfield{Name}_b$.\\
                                If no $\psi_i \in \Psi$ matches, Bob is not a recipient of $\mathcal{E}$, \textit{Exit}.
    \item\label{e:rem:check} Optionally check, if Alice tries to remove herself from the \ecf, i.e., Alice and Bob are identical.\\If $\publickey{B}{S} \overset{?}{=} \publickey{A}{S} = \convertskpk\left(\privatekey{A}{S}\right)$, \textit{May exit}.
    \item\label{e:rem:rem} Remove Bob: $\Psi' = \Psi \setminus \left\{\psi_b\right\}$.
    \item\label{e:rem:enc} Encrypt the modified \ecf: $\mathcal{E}' \leftarrow \encrypt^\textnormal{ECF}\left(\Psi', \Omega\right)$.
\end{enumerate}
As noted in \Cref{sec:ops:create}, the final \Cref{e:rem:enc} does not require any private keys for the encryption procedure. This implies that recipients of an \ecf\ are able to remove themselves. However, this is generally an undesired feature and, therefore, the implementation should prevent such action by executing the optional \Cref{e:rem:check}. Our \ac{PoC} implementation of the command line tool does not allow any recipient self-removal.

Finally, it must be noted that the restriction explained in \Cref{sec:app:remove} still holds: All former recipients are always able to access old versions of an \ecf. As stated earlier, this is unpreventable and should be mitigated by using short-lived secrets and by applying the principle of minimal privilege.

\subsection{Changing the Content}\label{sec:ops:content}
Modifying the confidential data stored within an \ecf\ is similar to modifying the recipient set. Following the workflow described in \Cref{sec:app:content}, we describe how the recipient Charlie can change the content of an \ecf. We denote the resulting bit string after modification as $\Omega' = \modify\left(\Omega\right)$ and Charlie's private key for signing as $\privatekey{C}{S}$:
\begin{enumerate}[(1)]
    \setlength{\itemsep}{4pt} 
    \item\label{e:mod:dec} Decrypt the \ecf: $\left(\Psi, \Omega\right) = \decrypt^\textnormal{ECF}\left(\privatekey{C}{S}, \mathcal{E}\right)$.
    \item\label{e:mod:mod} Modify the confidential data: $\Omega' = \modify\left(\Omega\right)$.
    \item\label{e:mod:enc} Encrypt the modified \ecf: $\mathcal{E}' \leftarrow \encrypt^\textnormal{ECF}\left(\Psi, \Omega'\right)$.
\end{enumerate}
Although possible, it is uncommon to change both the recipient set~$\Psi$ and the content~$\Omega$ in a single operation. This is why our \ac{PoC} command line tool offers both actions only separately.

\subsection{Generating $m - n$ Deception Blocks}\label{sec:ops:deception}
In order to achieve the goal of obfuscating the information about recipients in the public header of an \ecf, we take two measures. First, the \ecf\ format stores the names of recipients in the encrypted private body, thus, hiding them from external parties. Furthermore, if the public keys of users are not published to the open public, i.e., only visible to members of the same institution or company, external parties cannot determine if a given user is a recipient of an \ecf. This is because without knowledge of the public key, one cannot compute the $\textnormal{id\_tag}$ (cf.~\Cref{e:dec:id-tag} in~\Cref{sec:ops:use}) and therefore not search for a recipient-specific block within the public header of a given \ecf.

Second, we opted to obfuscate the true number of recipients $n$ in the public header. This is why $m \ge n$ gets chosen randomly in \Cref{e:enc:choose-m} in~\Cref{sec:ops:create}. Our \ac{PoC} implementation chooses $m$ randomly depending on $n$, such that $\max\left\{8, 2n\right\} \ge m \ge n$. After generating $n$ valid recipient-specific blocks in \Cref{e:enc:gen-rdi} in the encryption procedure, $m - n$ deception blocks must be generated to fill the public header and obtain a valid \ecf.

The main purpose of these deception blocks is to obfuscate~$n$. Therefore, these blocks must be generated in a way, such that an external party is not able to distinguish between real blocks and deception blocks. This is why the deception blocks should not be random bit strings because there is a possibility that the outputs of the used cryptographic algorithms used to generate the real blocks suffer from statistical biases.

To avoid distinguishability, we suggest a generation procedure as follows:
\begin{enumerate}[(1)]
    \setlength{\itemsep}{4pt} 
    \item\label{e:enc:gen-fake-full:keygen} Generate a new deception key pair:\\
                                            $\left(\privatekey{$\phi$}{S}, \publickey{$\phi$}{S}\right) \leftarrow \gen^\textnormal{S}, \quad\publickey{$\phi$}{X} = \convertSX\left(\publickey{$\phi$}{S}\right)$.
    \item\label{e:enc:gen-fake-full:id-tag} Compute \field{Identification Tag}\\$\textnormal{id\_tag} = \hash\left(\publickey{$\phi$}{S} \Vert \theta\right)\trunc{0}{15}$.
    \item\label{e:enc:gen-fake-full:gen-eph} Generate an ephemeral key pair for key exchange\\$\left(\privatekey{e}{X}, \publickey{e}{X}\right) \leftarrow \gen^\textnormal{X}$.
    \item\label{e:enc:gen-fake-full:gen-sym} Generate a random symmetric key $\symmetrickey{$\phi$} \leftarrow \gen^\textnormal{SYM}$.
    \item\label{e:enc:gen-fake-full:kex} Execute the key exchange algorithm:\\$\sharedsecret = \keyexchange\left(\privatekey{e}{X}, \publickey{$\phi$}{X}\right)$.
    \item\label{e:enc:gen-fake-full:pre2} Calculate $\symmetrickey{pre2} = \hash\left(\sharedsecret \Vert \publickey{$\phi$}{X} \Vert \publickey{e}{X}\right)\trunc{0}{y-1}$.
    \item\label{e:enc:gen-fake-full:pre1} Calculate $\symmetrickey{pre1} = \symmetrickey{$\phi$} \xor \symmetrickey{pre2}$.
    \item\label{e:enc:gen-fake-full:fin} Add $r_\phi = \textnormal{id\_tag} \Vert \publickey{e}{X} \Vert \symmetrickey{pre1}$ to $R$.
\end{enumerate}
However, generating deception blocks using this procedure is computationally more expensive than generating real blocks. In order to simplify the procedure, decrease the computational load, and therefore increase runtime performance, our \ac{PoC} implementation uses a different one. This is only possible if the used cryptographic hash function generates truly random looking bit strings, i.e., the outputs of \Cref{e:enc:gen-fake-full:id-tag,,e:enc:gen-fake-full:pre2} could have been generated at random instead. This implies that the output of \Cref{e:enc:gen-fake-full:pre1} also could have been generated at random, thus, eliminating most of the computationally expensive parts of the procedure. The assumption regarding the hash function is justified considering the lengths of the inputs, cf.~\autocite{CSE2020}. The simplified version of the deception block generation procedure reads as follows:
\begin{enumerate}[(1)]
    \setlength{\itemsep}{4pt} 
    \item\label{e:enc:gen-fake-fast:gen-eph} Generate an ephemeral key pair for key exchange\\$\left(\privatekey{e}{X}, \publickey{e}{X}\right) \leftarrow \gen^\textnormal{X}$.
    \item\label{e:enc:gen-fake-fast:id-tag} Generate a random \field{Identification Tag}:\\$\textnormal{id\_tag} \leftarrow \random\left(16\right)$.
    \item\label{e:enc:gen-fake-fast:pre1} Generate a random $\symmetrickey{pre1} \leftarrow \random\left(y\right)$.
    \item\label{e:enc:gen-fake-fast:fin} Add $r_\phi = \textnormal{id\_tag} \Vert \publickey{e}{X} \Vert \symmetrickey{pre1}$ to $R$.
\end{enumerate}
Of course, the generation of deception blocks does take some time, but the overhead is in the range of milliseconds as shown in \Cref{sec:improvements:perf}.

\begin{figure*}[t]
    \centering%
    \vspace*{-2mm}%
    \includegraphics[width=\textwidth]{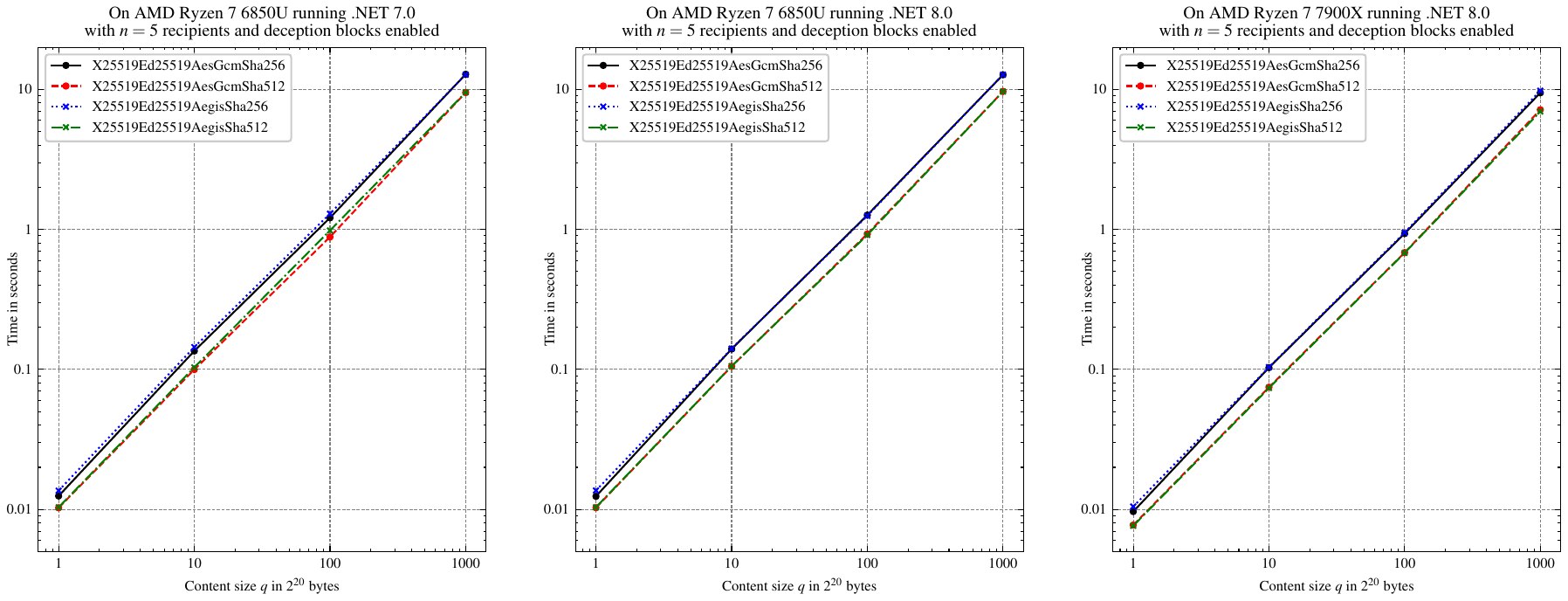}%
    \vspace*{-1mm}%
    \caption{Average encryption time for different content sizes, cipher suites, and runs.}%
    \label{fig:perf:encryption}%
    \vspace*{-3mm}%
\end{figure*}
\begin{figure*}[b]
    \centering%
    \vspace*{-1mm}%
    \includegraphics[width=\textwidth]{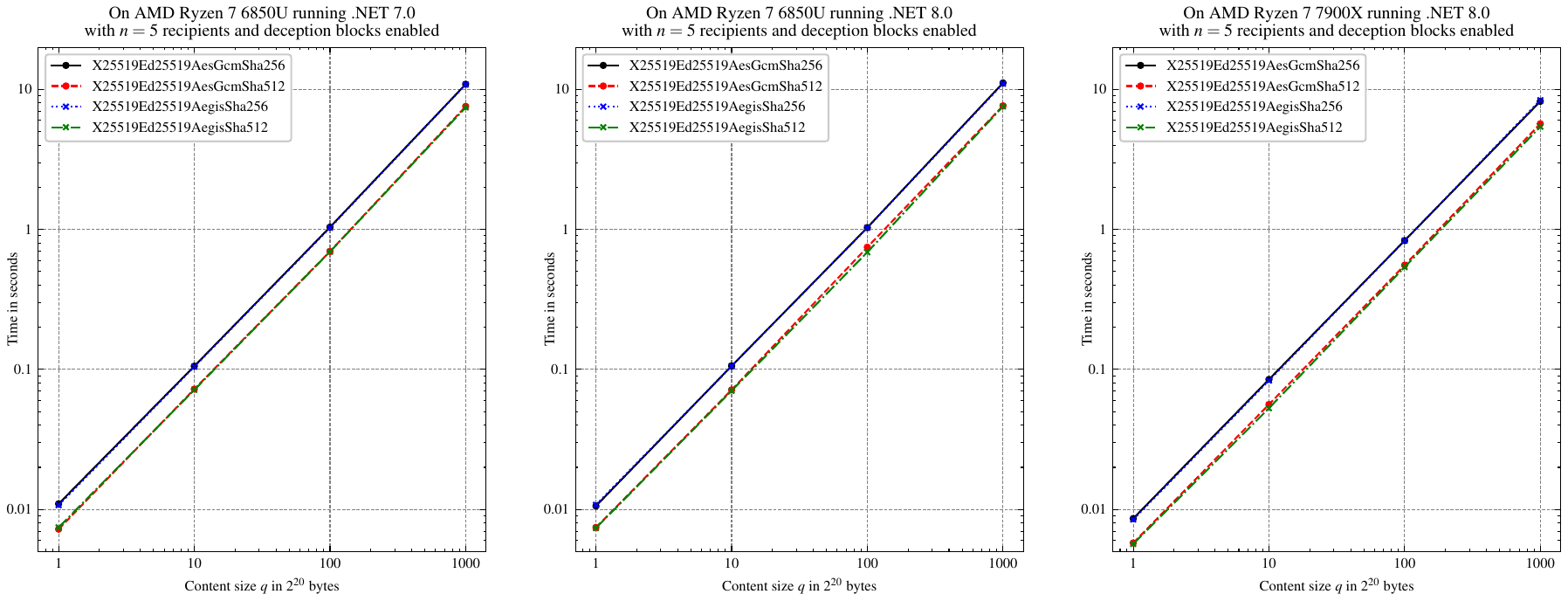}%
    \vspace*{-1mm}%
    \caption{Average decryption time for different content sizes, cipher suites, and runs.}%
    \label{fig:perf:decryption}%
    \vspace*{-2mm}%
\end{figure*}

\section{Recent Work and Improvements}\label{sec:improvements}
Since the original release of the \ecf\ format and the \ac{PoC} implementation in~\autocite{Bauer2023}, we extended and improved our work in different regards. The changes, enhancements, and new experiments are presented in the following subsections.

\subsection{Building on Linux and Dockerfile}\label{sec:improvements:build}
The original version of our \ac{PoC} implementation was targeted towards the Windows operating system. It could be cross-compiled for Linux, but building on Linux was not possible. However, enabling compilation on Linux was always a goal. The changes we made since the original release allow our \ac{PoC} implementation to be compiled on both operating systems and cross-compiled to the respective other operating system. Furthermore, using the \textit{single-file deployment}~\autocite{MicrosoftLearn2023} feature of recent .NET~\autocite{Microsoft2022} versions makes our \ac{PoC} implementation portable and easy to distribute without the need of installing further dependencies. Additionally to the build on Linux, we provide a \textit{Dockerfile} for compiling in a \textit{Docker}~\autocite{Docker2024} container.

Software and its dependencies need to be kept up-to-date, which is why we updated the targeted runtimes. This means that our \ac{PoC} implementation now can be compiled for \textit{.NET~7} and \textit{.NET~8}. We dropped support for \textit{.NET~6} in order to be able to use newer language features. The aforementioned \textit{Dockerfile} uses the most recent runtime. Next, we updated all libraries used for our \ac{PoC} implementation. This includes the \textit{Sodium} library~\autocite{Sodium2022}, which received new features, such as the support for the symmetric encryption scheme family AEGIS~\autocite{Denis2023}\autocite{Sodium2023aegis}. The wrapper library \textit{NSec}~\autocite{NSec2022} was also updated in order to be able to use the new algorithms in \textit{Sodium}.

\subsection{New Cipher Suite}\label{sec:improvements:cs}
As mentioned in \Cref{sec:requirements:structure}, we introduced two new cipher suites to the \ac{PoC} implementation. Namely, the symmetric encryption scheme \mbox{AEGIS-256}~\autocite{Denis2023}\autocite{Sodium2023aegis} was added to the set of cipher suites. Although the specifications for this new algorithm family are very recent and still \mbox{work-in-progress}, the authors of \textit{Sodium} recommend using \mbox{AEGIS-256} over \mbox{\ac{AES}-256-\ac{GCM}}~\autocite{Sodium2023aesgcm}. It is important to note that our use of \mbox{\ac{AES}-256-\ac{GCM}} never reaches the theoretical limit stated in~\autocite{Sodium2023aesgcm}, because the encryption procedure generates the symmetric encryption key every time anew (cf.~\Cref{sec:ops:create}).

Because of the changing nature of the AEGIS specification, we opted to keep the default cipher suite in our \ac{PoC} implementation. However, users of our command line tool may select the cipher suite that best suit their needs. Furthermore, our performance experiments in the following \Cref{sec:improvements:perf} hint that \Cref{e:cip:x25519-ed25519-aegis256-sha512} using AEGIS-256 as symmetric encryption scheme offers currently the best performance.

The flexibility of the \ecf\ format allows for different cipher suites to be implemented and the source code of our \ac{PoC} implementation is structured in a way to support the easy integration of more cryptographic algorithms. Therefore, implementing new cipher suites is straight-forward. This makes the \ecf\ format suitable in the long term since algorithms that are considered weak or are cracked in the future can be easily replaced by secure ones.\vspace*{-2mm}

\subsection{Performance Analysis}\label{sec:improvements:perf}
Extending our previous work~\autocite{Bauer2023}, we took performance measurements of our \ac{PoC} implementation on two different machines and in total using two different runtimes. The raw data of all runs is published alongside the code repository on GitHub. For the charts and tables in this paper, we aggregate the data by using average values. We provide the numeric values used to generate the performance charts in \Cref{sec:a:performance-tables}.

\begin{figure*}[t]
    \centering%
    \vspace*{-1mm}%
    \includegraphics[width=\textwidth]{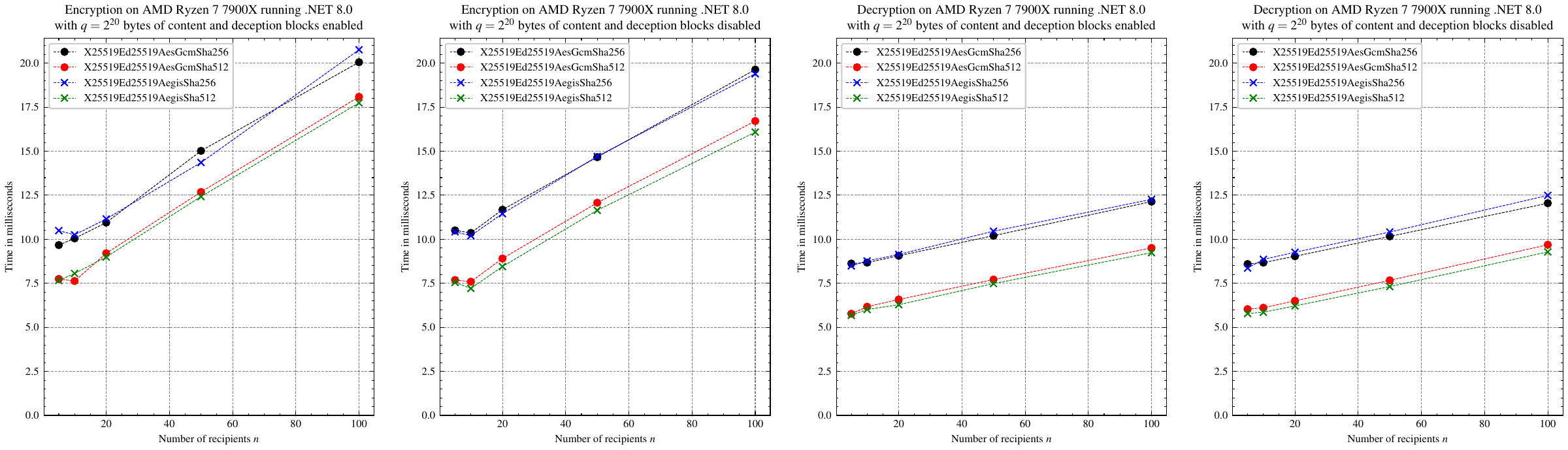}%
    \vspace*{-1mm}%
    \caption{Average encryption and decryption time for different number of recipients, cipher suites, and whether deception blocks are enabled.}%
    \label{fig:perf:enc-dec}%
    \vspace*{-5mm}%
\end{figure*}

The first experiment encrypts \ecfs\ with $n=5$ recipients and different content sizes ranging from \qtyrange{1}{1000}{\mebi\byte}. The results are depicted in \Cref{fig:perf:encryption}. We chose a log-log scale plot as both the content sizes and the amount of time it takes to encrypt the \ecf\ increase in orders of magnitudes. Although not directly visible in the plot, there is a linear relationship between the content size and the processing time, which can be validated using the numerical values in \Cref{tab:perf:content-size-enc} in \Cref{sec:a:performance-tables}. The differences between the runtimes \textit{.NET\,7.0} and \textit{.NET\,8.0} (left\,vs.\,middle chart) are negligible while a stronger processor (right\,chart) results in better absolute values.

Similarly, the chosen cipher suite makes a difference in performance. While the two symmetric encryption algorithms implemented in our \ac{PoC} perform almost identical, the hash function \mbox{SHA-512} (\Cref{e:cip:x25519-ed25519-aes256gcm-sha512,,e:cip:x25519-ed25519-aegis256-sha512}) outperforms its sibling \mbox{SHA-256} (\Cref{e:cip:x25519-ed25519-aes256gcm-sha256,,e:cip:x25519-ed25519-aegis256-sha256}) regardless of content size, number of recipients, runtime, and processor in all of our experiments. This is to be expected as the throughput of \mbox{SHA-512} is up to $1.6$ times~\autocite{CSE2015} as much as the throughput of \mbox{SHA-256}. Since our \ac{PoC} implementation uses the cryptographic library \textit{Sodium}~\autocite{Sodium2022}, which focuses on portability and therefore does not use any specialized processor instructions for \mbox{SHA-256}, our observed performance difference of about a factor of $1.5$ seems reasonable.

In a second experiment, we focus on the decryption speed and re-use the same parameters of $n=5$ and the content sizes of our previous experiment. \Cref{fig:perf:decryption} shows the results of this second experiment. We can see the same linear relationship between the content size and the processing time. The absolute decryption times are slightly lower than those for encryption, which must be due to the symmetric encryption and decryption algorithms. We can see the same speedup when using \mbox{SHA-512} over \mbox{SHA-256} as well as using a stronger processor and we observe little to no difference between the two symmetric encryption algorithms.

Our third experiment varies the number of recipients $n$ from \qtyrange{5}{1000}{} in order to make statements about the performance impact of generating the recipient-specific blocks in the public header. Furthermore, we can show that generating deception blocks increases the encryption time only slightly. We show the results in \Cref{fig:perf:enc-dec}, which are linear-linear plots and omit the samples for $n=1000$ for clarity. The numerical values are listed in \Cref{tab:perf:enc-dec} in \Cref{sec:a:performance-tables}. Deception blocks have no impact on the decryption, which is why the two decryption charts differ only in run-to-run variance. All four charts show a linear relationship between the number of recipients and the average processing time.

\begin{figure}[b]
    \centering%
    \vspace*{-5mm}%
    \includegraphics[width=\linewidth]{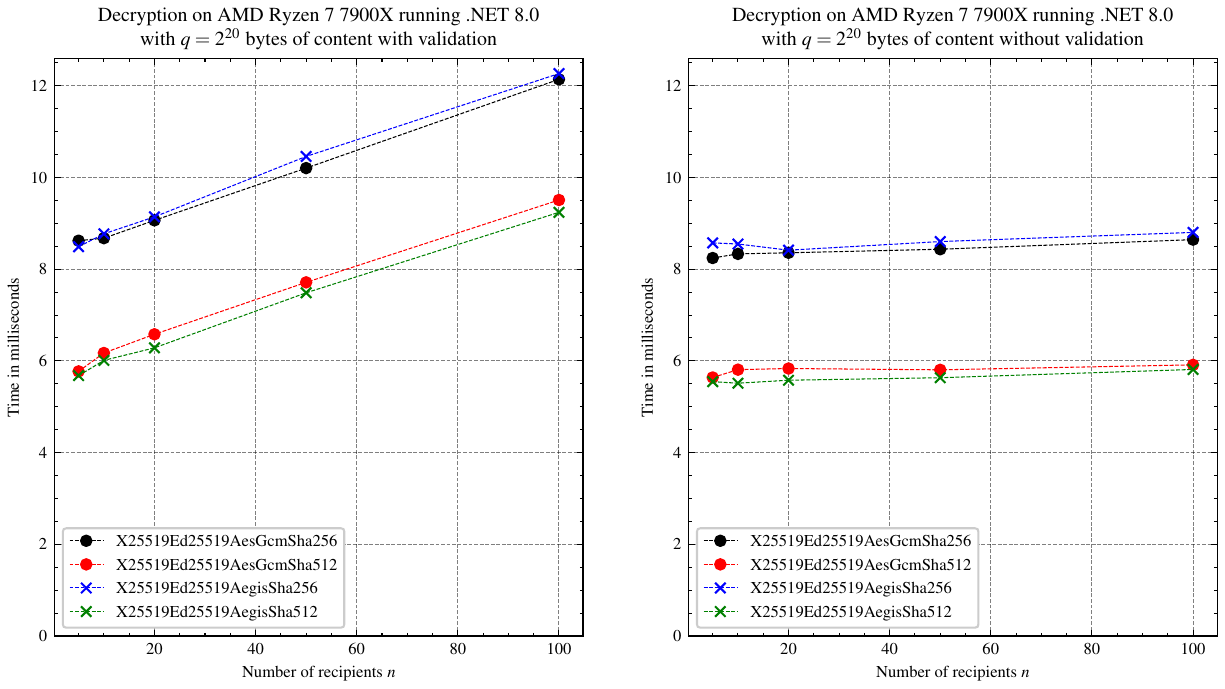}%
    \vspace*{-1mm}%
    \caption{Average decryption time for different number of recipients, cipher suites, and whether recipient signature validation is enabled.}%
    \label{fig:perf:dec-validation}%
    \vspace*{-5mm}%
\end{figure}

In a fourth experiment, we focused on the performance impact of validating all recipient signatures during decryption. Similar to the previous experiment, the number of recipients was varied while the content size was fixed to \qty{1}{\mebi\byte}. Activating the validation -- essentially executing \Cref{e:dec:verify-signatures} in \Cref{sec:ops:deception} -- results in a linear performance hit as depicted in \Cref{fig:perf:dec-validation}. Deactivating the validation step does not remove the linear dependency completely since the recipients still have to be parsed as stated in \Cref{sec:requirements:private}. It is, however, drastically decreased, offering a performance benefit to certain applicable situations described earlier. In general, we recommend keeping the recipient signature validation enabled.

Accomplishing our goal of performing the encryption and decryption procedure in less than \qty{100}{\milli\second} (cf.~\Cref{sec:requirements:goals}) is therefore dependent on the amount of confidential data stored within an \ecf\ as well as the number of recipients. Our experiments show that using a modern processor and choosing a cipher suite with \mbox{SHA-512} as the hash function can yield to execution times below \qty{100}{\milli\second} for content sizes smaller than \qty{10}{\mebi\byte}. As mentioned in \Cref{sec:app:remove}, an \ecf\ is designed to store passwords, certificate keys, and other similar credentials for a small set of recipients. The size of these types of content is typically in the range of $2^{5}$~to~$2^{15}$~bytes. Therefore, even the smallest content size in our experiments, $\qty{1}{\mebi\byte} = 2^{20}$~bytes, is more than sufficient for typical \ecf\ usage. Additionally, even though the \ecf\ format supports far more than $1000$ recipients, typically there are only a few recipients per file. Our experiments show that both the encryption and decryption procedure take less than \qty{15}{\milli\second} each with \qty{1}{\mebi\byte}-sized content and \qty{50}{} recipients. Therefore, we can safely state that the \ac{PoC} implementation fulfills the objective.

\subsection{Private Key Management}\label{sec:improvements:privatekey}
A major change was made to the management and storing of the private keys. While our original solution featured basic password-based encryption using a \ac{KDF}, we improved the implementation to support user-chosen parameters. Furthermore, the former static and predefined private key storage format was replaced by a more flexible one. This enables our \ac{PoC} implementation to be extended in the future with cipher suites using different Diffie-Hellman-like key exchange and matching signature algorithms. We discuss the option of choosing the key exchange algorithm independently from the signature algorithm in \Cref{sec:a:general-kex-sig}.

In order for the \ecf\ format to work, three components must be selected to match: the cipher suite, a private key, and the recipient information in the private body of an \ecf. This is why our \ac{PoC} implementation provides three base types \csharp{CipherSuite}, \csharp{ECFKey}, and \csharp{Recipient}. Then, the three subtypes \csharp{CSX25519Ed25519Base}, \csharp{EKX25519Ed25519}, and \csharp{RX25519Ed25519} are interlocked to form a group of matching components. Our implementation ensures this way that the correct type of key is used and the parsing of the recipient information blocks (cf.~\Cref{fig:private-recipient-block}) is done correctly.

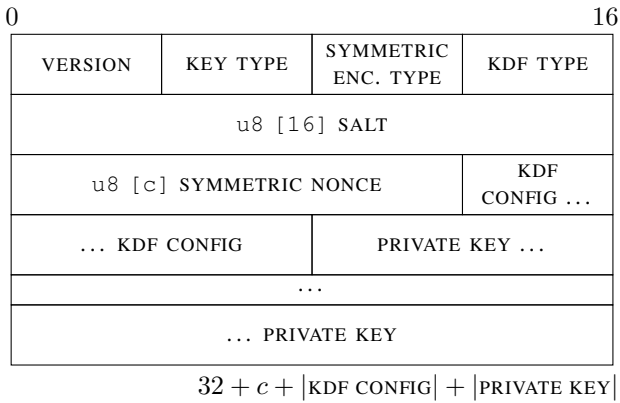
\begin{figure}[htb]
    \centering%
    \vspace*{-2mm}%
    \begin{tikzpicture}
        \node (version) at (0, 0) [u32le] {\field{Version}};
        \node (keytype) [right=0cm of version] [u32le] {\field{Key Type}};
        \node (symenc) [right=0cm of keytype] [u32le] {\field{Symmetric Enc. Type}};
        \node (kdftype) [right=0cm of symenc] [u32le] {\field{KDF Type}};
        \node (salt) [below right=0cm of version.south west] [u8a={16}] {\dt{u8 [16]}~\field{Salt}};
        \node (nonce) [below right=0cm of salt.south west] [u8a={12}] {\dt{u8 [c]}~\field{Symmetric Nonce}};
        \node (conf1) [right=0cm of nonce] [u8a={4}] {\field{KDF Config}~\dots};
        \node (conf2) [below right=0cm of nonce.south west] [u8a={8}] {\dots~\field{KDF Config}};
        \node (key1) [right=0cm of conf2] [u8a={8}] {\field{Private Key}~\dots};
        \node (key2) [below right=0cm of conf2.south west] [u8a={16}, half height] {\dots};
        \node (key3) [below=0cm of key2] [u8a={16}] {\dots~\field{Private Key}};

        \node [above right=0cm and \bitshoffset cm of version.north west] {$0$};
        \node [above left=0cm and \bitshoffset cm of kdftype.north east] {$16$};
        \node [below left=0cm and \bitshoffset cm of key3.south east] {$32 + c + \left|\field{KDF Config}\right| + \left|\field{Private Key}\right|$};
    \end{tikzpicture}
    \caption{Private key storage format.}\label{fig:private-key-storage-format}%
    \vspace*{-2mm}%
\end{figure}

The newly added, flexible format for private key storage is depicted in \Cref{fig:private-key-storage-format}. It starts similar to a public header with a \field{Version} to ensure future extensibility. The next three fields, \field{Key Type}, \field{Symmetric Encryption Type}, and \field{KDF Type} define which type of private key is stored, which symmetric encryption scheme was used to encrypt it, and which \ac{KDF} was used to derive the symmetric encryption key from the user-chosen password. Next, a \field{Salt} that is used for key derivation is stored, followed by a \field{Symmetric Nonce} used for encrypting and decrypting the private key.

Depending on the \ac{KDF}, its configuration is stored next in the field \field{KDF Config}. For the default algorithm Argon2id~\autocite{Biryukov2017}, three values are stored: the number of iterations, the memory size, and the degree of parallelism. These parameters can be chosen freely by the user as long as they meet the specification. We provide sensible defaults in our \ac{PoC} command line tool.

Finally, the encrypted private key is appended last. Its length is dependent on the cipher suite and the chosen symmetric encryption algorithm. All implemented cipher suites use an Ed25519 private key as their basis, since it can be converted into an Ed25519 public key and furthermore into both a X25519 private key and a X25519 public key. Therefore, it is only necessary to store the Ed25519 private key. However, when using different cipher suites (cf.~\Cref{sec:a:general-kex-sig}), it might be necessary to store multiple keys, which is why the storage format is designed as flexible as possible.

To keep the private key storage format simple, we opted to not include an encrypted hash value as a field. Instead, we utilize the \enquote{associated data} option of the \ac{AEAD}-based symmetric encryption algorithm. All unencrypted fields are used as input to the \enquote{associated data} option, which means the \ac{PoC} can detect modifications to any of the fields. This approach limits the choice of symmetric encryption schemes to \ac{AEAD}-based ones. However, since both symmetric encryption schemes used within our \ac{PoC} implementation are indeed \ac{AEAD}-based, there is actually no restriction. The user can choose which symmetric encryption scheme and which \ac{KDF} to use in order to protect the private key. Our \ac{PoC} implementation uses \ac{AES}-256-\ac{GCM}~\autocite{McGrew2005}\autocite{Dworkin2007} and Argon2id~\autocite{Biryukov2017} by default.

\section{Conclusion and Future Work}\label{sec:conclusion}
In this paper, we extended our previous work of \acl{ECF} in~\autocite{Bauer2023} and described the \ecf\ format in more detail. Additionally, we presented an example use case workflow in \Cref{sec:app}, which motivated the requirements, the design goals, and ultimately the \ecf\ format presented in \Cref{sec:requirements}. To demonstrate the \ecf\ format flexibility we claimed in our previous work, we implemented two new cipher suites in our \acl{PoC} implementation and focused on a cipher suite-independent presentation of the \ecf\ format and operations in \Cref{sec:ops}. Although all discussed cipher suites use a key exchange and a signature algorithm based on Curve25519~\autocite{Bernstein2006}~\autocite{Bernstein2012}, we provide guidance for implementing \ecf\ with different algorithms in \Cref{sec:a:general-kex-sig}.

\Cref{sec:improvements} describes more changes and improvements made to our \ac{PoC} implementation, which is now native to Linux and sports a new and improved private key management. We also took performance measurements in order to validate our claim of providing a fast-enough implementation to be used in production. We are confident that our work is ready to be used in production.

The full code of the updated and extended \ac{PoC} implementation as well as unit tests, performance tests, and performance analysis data for that code are available at:
\begin{center}
    \url{https://github.com/Hirnmoder/ECF}
\end{center}

The modular structure of our \ac{PoC} implementation allows for an easy extension and the implementation of more features and functionalities. We are looking forward to the feedback from the community to further improve the \ecf\ format and its implementation.

Future work may focus on finding further use cases and applications of \ecf. For example, the subject of applying \ecf\ or \ecf-based solutions in diverse contexts, such as within public authorities, raises interesting research questions and may lead to fruitful improvements of the \ecf\ format as well as our \ac{PoC} implementation.

Another area of work could be the implementation of a \ac{PKI}-based trust model amongst the recipients of an \ecf. Such extension could help establishing common trust anchors in multi-organization settings. In addition to that, changes to the \ecf\ format could be made to prove content modification ownership by employing digital signatures.

Furthermore, it may be possible to design a two-staged access control framework which allows a set of recipients to read the confidential data, but only a subset of those recipients to alter it. Such amendment could help distinguishing between different roles of entities accessing an \ecf.

~\par
Finally, the authors would like to thank the International Academy, Research, and Industry Association for the opportunity to present the idea of \aclp{ECF} in more detail, supplemented by current developments of the project.

\newpage
\enlargethispage{-2cm}
\advance\voffset by 1.5cm
\vspace*{-10.5mm}\printbibliography

\newpage
\advance\voffset by -1.5cm
\appendix

\subsection{Choosing the Key Exchange Algorithm Independently from the Signature Algorithm}\label{sec:a:general-kex-sig}
The \ecf\ format as well as our \ac{PoC} implementation use a well-defined conversion method to convert an Ed25519 private or public key to its corresponding X25519 private or public key~\autocite{CSE2019}\autocite{Sodium2022ed2c}, respectively. This implies that storing, using, and having access to the Ed25519-version of a private or public key is enough to carry out all signing and key agreement operations described in this paper. In order to save memory and remove unnecessary redundancy, we opted to employ this conversion method throughout the \ecf\ format and operations.

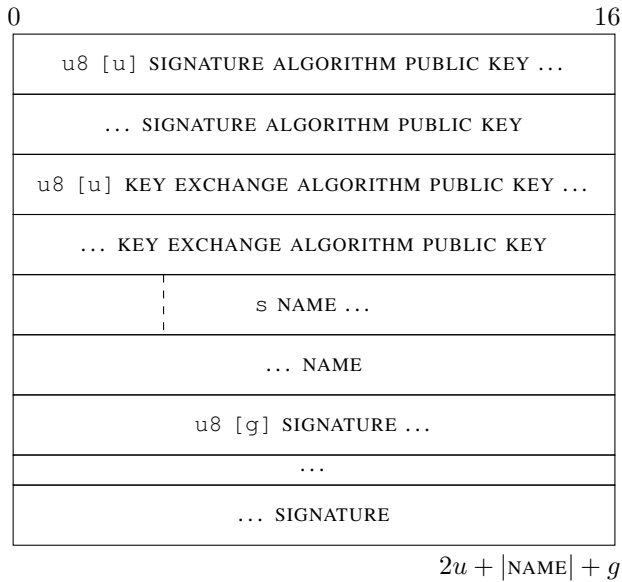
\begin{figure}[htbp]
    \centering
    \begin{tikzpicture}
        \node (pk1) at (0, 0) [u8a={16}] {\dt{u8 [u]} \field{Signature Algorithm Public Key}~\dots};
        \node (pk2) [below=0cm of pk1] [u8a={16}] {\dots~\field{Signature Algorithm Public Key}};
        \node (pk3) [below=0cm of pk2] [u8a={16}] {\dt{u8 [u]} \field{Key Exchange Algorithm Public Key}~\dots};
        \node (pk4) [below=0cm of pk3] [u8a={16}] {\dots~\field{Key Exchange Algorithm Public Key}};
        \node (name1) [below=0cm of pk4] [u8a={16}] {\dt{s} \field{Name}~\dots};
        \draw [dashed] ($(name1.north west) + (\uintwidth cm, 0)$) -- +(0, -\blockheight cm);
        \node (name2) [below=0cm of name1] [u8a={16}] {\dots~\field{Name}};
        \node (sig1) [below=0cm of name2] [u8a={16}] {\dt{u8 [g]} \field{Signature}~\dots};
        \node (sig2) [below=0cm of sig1] [u8a={16}, half height] {\dots};
        \node (sig3) [below=0cm of sig2] [u8a={16}] {\dots~\field{Signature}};

        \node [above right=0cm and \bitshoffset cm of pk1.north west] {$0$};
        \node [above left=0cm and \bitshoffset cm of pk1.north east] {$16$};
        \node [below left=0cm and \bitshoffset cm of sig3.south east] {$2u + \left|\field{Name}\right| + g$};
    \end{tikzpicture}
    \caption{Recipient information block structure within the private body of an \ecf\ when no key conversion method is available to convert between the keys of the key exchange and the signature algorithm.}
    \label{fig:private-recipient-block-kex-sig}
\end{figure}

However, different key exchange and signature algorithms may not support such key conversion. In this case, the recipient information block stored inside the private body of an \ecf\ (cf.~\Cref{fig:private-recipient-block}) must include not one but two public keys: one for the signature algorithm and one for the key exchange algorithm. We depict the altered structure in \Cref{fig:private-recipient-block-kex-sig} and assume for simplicity reasons that both public keys have the same length of $u$~bytes. This is, however, not a requirement.

The field \field{Signature} now contains a signature not only over the \field{Name}, but over the combination of \field{Key Exchange Algorithm Public Key} and \field{Name}. This is compulsory in order to create a strong connection between the public key for key exchange and the human-readable name of a recipient.

The presented changes in data structure also need to be reflected in the \ecf\ operations. The recipient entry is now constructed as follows:
\begin{align*}
    \psi &= \publickey{$\psi$}{S} \Vert \publickey{$\psi$}{X} \Vert \mfield{Name} \Vert t \\
    t &= \sign\left(\privatekey{$\psi$}{S}, \publickey{$\psi$}{X} \Vert \utf\left(\mfield{Name}\right)\right)
\end{align*}

During encryption, $\publickey{$\psi_i$}{X}$ is already present and, therefore, \Cref{e:enc:gen-rdi:cvt} in \Cref{sec:ops:create} is omitted. In \Cref{e:enc:write-private}, the extended recipient information block structure is used.

The decryption procedure (cf.~\Cref{sec:ops:use}) is extended and now needs three input parameters:
\[ \left(\Psi, \Omega\right) = \decrypt^\textnormal{ECF} \left(\privatekey{}{S}, \privatekey{}{X}, \mathcal{E}\right) \]
Similarly to the encryption procedure, the first conversion in \Cref{e:dec:cvt} in \Cref{sec:ops:use} is omitted. Next, the deconstruction in \Cref{e:dec:verify-signatures:load} produces four values $\publickey{$\psi_i$}{S}$, $\publickey{$\psi_i$}{X}$, $\mfield{Name}_i$, and $t_i$ and the verification in \Cref{e:dec:verify-signatures:verify} needs to be adjusted, too:
\[ \textnormal{If~} \verify^\textnormal{S}\left(\publickey{$\psi_i$}{S}, \publickey{$\psi_i$}{X} \Vert \utf\left(\mfield{Name}_i\right), t_i\right) \overset{?}{=} 0 \textnormal{, \textit{Exit}.} \]

Adding new recipients to an \ecf\ requires them to generate their recipient entries as shown in \Cref{sec:ops:recipients}. The computations for $t_b$ and $\psi_b$ need to be adjusted accordingly and include the public key for key exchange as described above. This is also true for the verification in \Cref{e:add:verify} in \Cref{sec:ops:recipients}. The mandatory check in \Cref{e:add:check-pubk} now needs to check for both $\publickey{B}{S}$ and $\publickey{B}{X}$.

The optimized procedure for generating deception blocks, which is described in \Cref{sec:ops:deception}, does not require any modifications. This is because the \field{Identification Tag} is generated randomly instead of being derived from a key pair and, thus, no key conversion is necessary.

\subsection{Variables and Functions}\label{sec:a:variables}
\Cref{tab:variables} contains all simple variables used primarily to describe the \ecf\ format. The encryption and decryption procedures make use of various cryptographic algorithms and their respective input and output parameters. These are defined and described in \Cref{tab:component-variables}. All used functions are listed in \Cref{tab:functions}. A randomized or probabilistic algorithm is indicated by the left arrow \enquote{$\leftarrow$}, whereas an equal sign \enquote{$=$} indicates deterministic execution.

\begin{table}[htb]
    \centering%
    \renewcommand{\arraystretch}{1.5}%
    \caption{Variables.}\label{tab:variables}%
    \vspace*{-2mm}%
    \begin{NiceTabular}{>{\raggedleft}p{0.65cm}@{}>{$}l<{$}@{\hspace{2mm}}p{\linewidth - 2cm}}
        \multicolumn{2}{r}{Variable} & Description \\\hline
        & a  & length of \field{Key Agreement Information} in bytes\\
        & b  & length of the encrypted private body in bytes\\
        & b' & length of the unencrypted private body in bytes\\
        & c  & length of \field{Symmetric Nonce} in bytes\\
        & d  & length of the output of the hash function in bytes\\
        & g  & length of a signature in bytes\\
        & h  & length of the public header in bytes\\
        & m  & number of recipients in the public header\\
        & n  & number of true recipients in the private body\\
        & o_i& offset of recipient-specific block $i$ within the file in bytes\\
        & q  & length of \field{Content} in bytes\\
        & \bar{q} & length of all private body fields except \field{Content} in bytes\\
        & u  & length of a public key in bytes\\
        & v  & length of a private key in bytes\\
        & y  & length of the symmetric key in bytes\\
    \end{NiceTabular}
    \vspace*{-4mm}%
\end{table}

\begin{table}[!t]
    \centering%
    \vspace*{2mm}%
    \renewcommand{\arraystretch}{1.5}%
    \caption{Component Variables.}\label{tab:component-variables}%
    \vspace*{-2mm}%
    \begin{NiceTabular}{>{$}m{2.7cm}<{$}@{\hspace{2mm}}m{\linewidth - 3.5cm}}
        \textnormal{Variable} & Description \\\hline
        \publickey{A}{X} \in \left\{0, 1\right\}^{8u} & \underline{p}ublic \underline{k}ey of \underline{A}lice for key e\underline{x}change \\
        \privatekey{B}{X} \in \left\{0, 1\right\}^{8v} & private/\underline{s}ecret \underline{k}ey of \underline{B}ob for key e\underline{x}change \\
        \symmetrickey{final} \in \left\{0, 1\right\}^{8y} & final symmetric encryption \underline{k}ey \\
        \alpha \in \left\{0, 1\right\}^{8c} & symmetric nonce \\
        \theta \in \left\{0, 1\right\}^{8\cdot16} & salt \\
        t \in \left\{0, 1\right\}^{8g} & a signature $t$ is $g$~bytes long \\
        \beta \in \left\{0, 1\right\}^{8d} & hash function's output is always $d$ bytes \\
        r = \textnormal{id\_tag} \Vert \publickey{e}{X} \Vert \symmetrickey{pre1} & \underline{r}ecipient in public header \\
        R = \left\{r_0, \dots, r_{m-1}\right\} & recipient set in public header \\
        \psi = \publickey{$\psi$}{S} \Vert \mfield{Name} \Vert t & recipient in private body \\
        \Psi = \left\{\psi_0, \dots, \psi_{n-1} \right\} & recipient set in private body \\
        \Omega \in \left\{0, 1\right\}^{8q} & confidential data stored in an \ecf, $q$~bytes long \\
        \mathcal{E} = \mathcal{H} \Vert \mathcal{B} \Vert \beta_\textnormal{all} & an \ecf\ consists of a public header~$\mathcal{H}$, a private body~$\mathcal{B}$, and a public footer $\beta_\textnormal{all}$
    \end{NiceTabular}
\end{table}

\begin{table}[!b]
    \centering%
    \vspace*{-2mm}%
    \renewcommand{\arraystretch}{1.5}%
    \caption{Functions.}\label{tab:functions}%
    \vspace*{-2mm}%
    \begin{NiceTabular}{>{$}m{3.35cm}<{$}@{\hspace{1mm}}m{\linewidth - 3.86cm}}
        \textnormal{Function Definition} & Description \\\hline
        x = x_1 \Vert x_2 & concatenation of two bit strings \\
        x = x_1 \xor x_2 & bitwise exclusive OR (XOR) operation on two same-length bit strings \\
        x' = x\trunc{0}{j-1} & truncation of the bit string $x \in \left\{0, 1\right\}^{l \ge 8j}$ to the first $j$ bytes \\

        \left(\privatekey{}{S}, \publickey{}{S}\right) \leftarrow \gen^\textnormal{S} & \underline{gen}erates a key pair for \underline{s}igning \\
        \left(\privatekey{}{X}, \publickey{}{X}\right) \leftarrow \gen^\textnormal{X} & \underline{gen}erates a key pair for key e\underline{x}change \\
        \parbox{3.2cm}{$\privatekey{}{X} = \convertSX\left(\privatekey{}{S}\right)$\\$\publickey{}{X} = \convertSX\left(\publickey{}{S}\right)$} & converts a private/public key for signing to a private/public key for key exchange $^{\left(*\right)}$ \\[3mm]
        \parbox{3.35cm}{$\publickey{}{S} = \convertskpk\left(\privatekey{}{S}\right)$\\$\publickey{}{X} = \convertskpk\left(\privatekey{}{X}\right)$} & converts a private key to a public key \\[2mm]
        
        \symmetrickey{} \leftarrow \gen^\textnormal{SYM} & \underline{gen}erates a \underline{sym}metric encryption key \\
        x = \utf\left(s\right) & converts the character string $s$ without BOM into the UTF-8 bit string $x$ \\
        x = \random\left(j\right) & generates a random $j$-byte long bit string~$x$ \\
        
        \parbox{3cm}{$\sharedsecret = \keyexchange\left(\privatekey{A}{X}, \publickey{B}{X}\right)$\\[1mm]\hspace*{\widthof{$\sharedsecret~$}}$= \keyexchange\left(\privatekey{B}{X}, \publickey{A}{X}\right)$} & obtain a \underline{s}hared \underline{s}ecret by performing the \underline{k}ey \underline{ex}change with a secret and a public key \\
        t \leftarrow \sign\left(\privatekey{A}{S}, x\right) &  \underline{sign}ing the bit string $x$ with \underline{A}lice's private key for \underline{s}igning yields a signature $t$ \\
        \verify^\textnormal{S}\left(\publickey{A}{S}, x, t\right) \in \left\{0, 1\right\} & returns $1$, if and only if the signature $t$ is valid for the given public key and the bit string $x$, otherwise $0$ \\
        \gamma \leftarrow \encrypt^\textnormal{SYM}\left(\symmetrickey{}, \alpha, x\right) & \underline{enc}rypting the bit string $x$ with a \underline{sym}metric encryption key $\symmetrickey{}$ and a nonce~$\alpha$ produces a ciphertext bit string $\gamma$ \\
        x' = \decrypt^\textnormal{SYM}\left(\symmetrickey{}, \alpha, \gamma\right) & \underline{dec}rypting a ciphertext bit string $\gamma$ with the correct \underline{sym}metric encryption key $\symmetrickey{}$ and the correct nonce $\alpha$ produces the bit string $x'$ \\
        \beta = \hash\left(x\right) & hash functions take any-length bit strings $x$ and produce a fixed-sized output $\beta$ \\
        \verify^\textnormal{H}\left(x, \beta\right) \in \left\{0, 1\right\} & returns $1$, if and only if the hash value $\beta$ matches the hash value of the bit string $x$, otherwise $0$ \\[5mm]\hline
        \multicolumn{2}{p{\linewidth - 1cm}}{\hspace*{-5mm}\footnotesize$^{\left(*\right)}$ This behavior is specific for Ed25519~\autocite{Bernstein2012} and X25519~\autocite{Bernstein2006} using the conversion method described in \autocite{CSE2019}\autocite{Sodium2022ed2c}. However, if a cipher suite does not allow such conversion, cf.~\Cref{sec:a:general-kex-sig}.}
    \end{NiceTabular}
    \vspace*{-5mm}
\end{table}

\subsection{Performance Analysis Value Tables}\label{sec:a:performance-tables}
This subsection provides the numerical values of the performance charts in \Cref{sec:improvements:perf}. All tables contain the aggregated data used to draw the performance charts. However, \Cref{tab:perf:enc-dec,,tab:perf:dec-validation} contain additional data points for $n=1000$ recipients, showing that our \ac{PoC} implementation is capable of handling even such unrealistic parameters efficiently. The raw data and the code to aggregate and plot the performance measurements is published on GitHub.

\begin{table}[htb]
    \renewcommand{\arraystretch}{1.3}%
    \centering%
    \caption{Average encryption time (in seconds) for different content sizes and cipher suites. Visualized in \Cref{fig:perf:encryption}.}\label{tab:perf:content-size-enc}%
    \begin{tabular}{r@{\hspace{2mm}}c@{\hspace{3mm}}rrrr} 
    \multirow[c]{2}{1cm}{\raggedleft Content Size} & \multirow[c]{2}{*}{Run}  & \multicolumn{4}{c}{Cipher Suite} \\
    & & \multicolumn{1}{c}{\ref{e:cip:x25519-ed25519-aes256gcm-sha256}} & \multicolumn{1}{c}{\ref{e:cip:x25519-ed25519-aes256gcm-sha512}} & \multicolumn{1}{c}{\ref{e:cip:x25519-ed25519-aegis256-sha256}} & \multicolumn{1}{c}{\ref{e:cip:x25519-ed25519-aegis256-sha512}} \\
    \hline\\[-3mm]
    \multirow[c]{3}{*}{\qty{1}{\mebi\byte}} & R7 6850U (.NET 7.0) & 0.012 & 0.010 & 0.014 & 0.010 \\
     & R7 6850U (.NET 8.0) & 0.012 & 0.010 & 0.014 & 0.010 \\
     & R7 7900X (.NET 8.0) & 0.010 & 0.008 & 0.010 & 0.008 \\
    \\[-3mm]\cline{1-6}\\[-3mm]
    \multirow[c]{3}{*}{\qty{10}{\mebi\byte}} & R7 6850U (.NET 7.0) & 0.136 & 0.100 & 0.144 & 0.103 \\
     & R7 6850U (.NET 8.0) & 0.140 & 0.105 & 0.141 & 0.106 \\
     & R7 7900X (.NET 8.0) & 0.103 & 0.075 & 0.104 & 0.074 \\
    \\[-3mm]\cline{1-6}\\[-3mm]
    \multirow[c]{3}{*}{\qty{100}{\mebi\byte}} & R7 6850U (.NET 7.0) & 1.210 & 0.883 & 1.295 & 0.983 \\
     & R7 6850U (.NET 8.0) & 1.265 & 0.928 & 1.250 & 0.913 \\
     & R7 7900X (.NET 8.0) & 0.934 & 0.683 & 0.950 & 0.680 \\
    \\[-3mm]\cline{1-6}\\[-3mm]
    \multirow[c]{3}{*}{\qty{1000}{\mebi\byte}} & R7 6850U (.NET 7.0) & 12.822 & 9.490 & 12.740 & 9.508 \\
     & R7 6850U (.NET 8.0) & 12.711 & 9.665 & 12.726 & 9.650 \\
     & R7 7900X (.NET 8.0) & 9.449 & 7.153 & 9.782 & 6.948 \\
    \end{tabular}
\end{table}

\begin{table}[htb]
    \renewcommand{\arraystretch}{1.3}%
    \centering%
    \caption{Average decryption time (in seconds) for different content sizes and cipher suites. Visualized in \Cref{fig:perf:decryption}.}\label{tab:perf:content-size-dec}%
    \begin{tabular}{r@{\hspace{2mm}}c@{\hspace{3mm}}rrrr} 
    \multirow[c]{2}{1cm}{\raggedleft Content Size $q$} & \multirow[c]{2}{*}{Run}  & \multicolumn{4}{c}{Cipher Suite} \\
    & & \multicolumn{1}{c}{\ref{e:cip:x25519-ed25519-aes256gcm-sha256}} & \multicolumn{1}{c}{\ref{e:cip:x25519-ed25519-aes256gcm-sha512}} & \multicolumn{1}{c}{\ref{e:cip:x25519-ed25519-aegis256-sha256}} & \multicolumn{1}{c}{\ref{e:cip:x25519-ed25519-aegis256-sha512}} \\
    \hline\\[-3mm]
    \multirow[c]{3}{*}{\qty{1}{\mebi\byte}} & R7 6850U (.NET 7.0) & 0.011 & 0.007 & 0.011 & 0.007 \\
     & R7 6850U (.NET 8.0) & 0.011 & 0.007 & 0.011 & 0.007 \\
     & R7 7900X (.NET 8.0) & 0.009 & 0.006 & 0.008 & 0.006 \\
    \\[-3mm]\cline{1-6}\\[-3mm]
    \multirow[c]{3}{*}{\qty{10}{\mebi\byte}} & R7 6850U (.NET 7.0) & 0.106 & 0.072 & 0.105 & 0.071 \\
     & R7 6850U (.NET 8.0) & 0.106 & 0.071 & 0.106 & 0.070 \\
     & R7 7900X (.NET 8.0) & 0.085 & 0.056 & 0.083 & 0.053 \\
    \\[-3mm]\cline{1-6}\\[-3mm]
    \multirow[c]{3}{*}{\qty{100}{\mebi\byte}} & R7 6850U (.NET 7.0) & 1.038 & 0.696 & 1.028 & 0.693 \\
     & R7 6850U (.NET 8.0) & 1.030 & 0.746 & 1.028 & 0.687 \\
     & R7 7900X (.NET 8.0) & 0.832 & 0.555 & 0.831 & 0.538 \\
    \\[-3mm]\cline{1-6}\\[-3mm]
    \multirow[c]{3}{*}{\qty{1000}{\mebi\byte}} & R7 6850U (.NET 7.0) & 10.882 & 7.557 & 10.836 & 7.436 \\
     & R7 6850U (.NET 8.0) & 11.113 & 7.626 & 10.992 & 7.559 \\
     & R7 7900X (.NET 8.0) & 8.212 & 5.677 & 8.366 & 5.415 \\
    \end{tabular}
\end{table}
\clearpage

\begin{table}[!t]
    \renewcommand{\arraystretch}{1.3}%
    \centering%
    \caption{Average encryption and decryption times (in milliseconds) for different number of recipients and cipher suites. Visualized in \Cref{fig:perf:enc-dec}.}\label{tab:perf:enc-dec}%
    \begin{tabular}{r@{\hspace{2mm}}c@{\hspace{2mm}}c@{\hspace{2mm}}rrrr} 
    \multirow[c]{2}{*}{\raggedleft $n$} & \multirow[c]{2}{*}{Opera\-tion} & Deception & \multicolumn{4}{c}{Cipher Suite} \\
    & & blocks & \multicolumn{1}{c}{\ref{e:cip:x25519-ed25519-aes256gcm-sha256}} & \multicolumn{1}{c}{\ref{e:cip:x25519-ed25519-aes256gcm-sha512}} & \multicolumn{1}{c}{\ref{e:cip:x25519-ed25519-aegis256-sha256}} & \multicolumn{1}{c}{\ref{e:cip:x25519-ed25519-aegis256-sha512}} \\
    \hline\\[-3mm]
    \multirow[c]{4}{*}{5} & \multirow[c]{2}{*}{$\encrypt$} & \yes & 9.671 & 7.745 & 10.493 & 7.659 \\
     &  & \no & 10.506 & 7.688 & 10.424 & 7.552 \\
    \cline{2-7}
     & \multirow[c]{2}{*}{$\decrypt$} & \yes & 8.622 & 5.771 & 8.490 & 5.679 \\
     &  & \no & 8.594 & 6.032 & 8.364 & 5.784 \\
    \\[-3mm]\cline{1-7}\\[-3mm]
    \multirow[c]{4}{*}{10} & \multirow[c]{2}{*}{$\encrypt$} & \yes & 10.050 & 7.630 & 10.253 & 8.061 \\
     &  & \no & 10.363 & 7.593 & 10.205 & 7.211 \\
    \cline{2-7}
     & \multirow[c]{2}{*}{$\decrypt$} & \yes & 8.676 & 6.172 & 8.771 & 6.011 \\
     &  & \no & 8.675 & 6.111 & 8.859 & 5.858 \\
    \\[-3mm]\cline{1-7}\\[-3mm]
    \multirow[c]{4}{*}{20} & \multirow[c]{2}{*}{$\encrypt$} & \yes & 10.948 & 9.204 & 11.154 & 8.992 \\
     &  & \no & 11.680 & 8.909 & 11.457 & 8.451 \\
    \cline{2-7}
     & \multirow[c]{2}{*}{$\decrypt$} & \yes & 9.065 & 6.580 & 9.141 & 6.285 \\
     &  & \no & 9.038 & 6.507 & 9.264 & 6.216 \\
    \\[-3mm]\cline{1-7}\\[-3mm]
    \multirow[c]{4}{*}{50} & \multirow[c]{2}{*}{$\encrypt$} & \yes & 15.023 & 12.688 & 14.358 & 12.424 \\
     &  & \no & 14.664 & 12.076 & 14.706 & 11.646 \\
    \cline{2-7}
     & \multirow[c]{2}{*}{$\decrypt$} & \yes & 10.204 & 7.713 & 10.459 & 7.485 \\
     &  & \no & 10.164 & 7.670 & 10.414 & 7.315 \\
    \\[-3mm]\cline{1-7}\\[-3mm]
    \multirow[c]{4}{*}{100} & \multirow[c]{2}{*}{$\encrypt$} & \yes & 20.055 & 18.090 & 20.763 & 17.747 \\
     &  & \no & 19.636 & 16.715 & 19.402 & 16.085 \\
    \cline{2-7}
     & \multirow[c]{2}{*}{$\decrypt$} & \yes & 12.141 & 9.509 & 12.266 & 9.240 \\
     &  & \no & 12.049 & 9.689 & 12.491 & 9.284 \\
    \\[-3mm]\cline{1-7}\\[-3mm]
    \multirow[c]{4}{*}{1000} & \multirow[c]{2}{*}{$\encrypt$} & \yes & 119.864 & 115.551 & 119.596 & 115.086 \\
     &  & \no & 104.953 & 101.514 & 106.809 & 99.056 \\
    \cline{2-7}
     & \multirow[c]{2}{*}{$\decrypt$} & \yes & 46.000 & 43.972 & 47.385 & 42.321 \\
     &  & \no & 46.109 & 43.981 & 47.173 & 42.594 \\
    \end{tabular}
\end{table}

~\pagebreak~

\begin{table}[!t]
    \vspace*{-15.45cm}
    \renewcommand{\arraystretch}{1.3}%
    \centering%
    \caption{Average decryption time (in milliseconds) for different number of recipients and whether recipient validation is enabled. Visualized in \Cref{fig:perf:dec-validation}.}\label{tab:perf:dec-validation}%
    \begin{tabular}{rcrrrr} 
    \multirow[c]{2}{*}{\raggedleft $n$} & \multirow[c]{2}{*}{Validation}  & \multicolumn{4}{c}{Cipher Suite} \\
    & & \multicolumn{1}{c}{\ref{e:cip:x25519-ed25519-aes256gcm-sha256}} & \multicolumn{1}{c}{\ref{e:cip:x25519-ed25519-aes256gcm-sha512}} & \multicolumn{1}{c}{\ref{e:cip:x25519-ed25519-aegis256-sha256}} & \multicolumn{1}{c}{\ref{e:cip:x25519-ed25519-aegis256-sha512}} \\
    \hline\\[-3mm]
    \multirow[c]{2}{*}{5} & \yes & 8.622 & 5.771 & 8.490 & 5.679 \\
     & \no & 8.244 & 5.639 & 8.573 & 5.547 \\
    \\[-3mm]\cline{1-6}\\[-3mm]
    \multirow[c]{2}{*}{10} & \yes & 8.676 & 6.172 & 8.771 & 6.011 \\
     & \no & 8.334 & 5.809 & 8.549 & 5.513 \\
    \\[-3mm]\cline{1-6}\\[-3mm]
    \multirow[c]{2}{*}{20} & \yes & 9.065 & 6.580 & 9.141 & 6.285 \\
     & \no & 8.357 & 5.833 & 8.414 & 5.578 \\
    \\[-3mm]\cline{1-6}\\[-3mm]
    \multirow[c]{2}{*}{50} & \yes & 10.204 & 7.713 & 10.459 & 7.485 \\
     & \no & 8.435 & 5.805 & 8.602 & 5.633 \\
    \\[-3mm]\cline{1-6}\\[-3mm]
    \multirow[c]{2}{*}{100} & \yes & 12.141 & 9.509 & 12.266 & 9.240 \\
     & \no & 8.644 & 5.914 & 8.802 & 5.813 \\
    \\[-3mm]\cline{1-6}\\[-3mm]
    \multirow[c]{2}{*}{1000} & \yes & 46.000 & 43.972 & 47.385 & 42.321 \\
     & \no & 11.239 & 8.300 & 11.366 & 7.840 \\
    \end{tabular}
\end{table}

\end{document}